\documentclass[journal=tches]{iacrtrans}
\usepackage[utf8]{inputenc}
\usepackage{tugraz_defaults}
\usepackage{setspace}

\usepackage{algorithm}
\makeatletter
\renewcommand{\ALG@name}{Pseudocode}
\makeatother

\usepackage[noend]{algpseudocode}
\usepackage{varwidth}

\newif\ifanonymous

\newcommand{\fipac}{\mbox{FIPAC}\xspace}

\newcommand{\speccoderange}{\mbox{54--97\,\%}\xspace}
\newcommand{\specruntimerange}{\mbox{35--105\,\%}\xspace}
\newcommand{\embenchruntimerange}{\mbox{61--211\,\%}\xspace}

\newcommand{\pacia}{\mbox{\texttt{PACIA}}\xspace}
\newcommand{\autiza}{\mbox{\texttt{AUTIZA}}\xspace}

\ifanonymous
\author{Anonymous author(s)}
\else
\author[R. Schilling, P. Nasahl, S. Mangard]{Robert Schilling \and Pascal Nasahl \and Stefan Mangard}

\institute{Graz University of Technology, Graz, Austria
\\\email{first.last@iaik.tugraz.at}
}
\fi

\title[\fipac: Thwarting Control-Flow Attacks with ARM Pointer Authentication]{\fipac: Thwarting Fault- and Software-Induced Control-Flow Attacks with ARM Pointer Authentication}

\begin{document}

\maketitle

\keywords[Control-Flow Integrity, Fault Attacks, ARM PAC]{Control-Flow Integrity \and Fault Attacks
  \and ARM PAC}

\begin{abstract}
With the improvements of computing technology, more and more applications in the Internet-of-Things, mobile devices, automotive, or industrial areas embed powerful ARM processors into their devices.
These systems can be attacked by redirecting the control-flow of a program to bypass critical pieces of code such as privilege checks or signature verifications or to perform other fault attacks on applications, operating systems, or security mechanisms like secure boot.
Control-flow hijacks can be performed using classical software vulnerabilities, physical fault attacks, or software-induced fault attacks, such as CLKscrew or Plundervolt.
To cope with this threat and to protect the control-flow, dedicated countermeasures are needed.

To counteract control-flow hijacks, control-flow integrity~(CFI) aims to be a generic solution.
However, software-based CFI protection schemes typically either protect against software or fault attacks, but not against both.
While hardware-assisted CFI schemes can mitigate both types of attacks, they require extensive hardware modifications.
As hardware changes are unrealistic for existing ARM architectures, a wide range of systems remains unprotected and vulnerable to control-flow attacks.



In this work, we present \fipac, an efficient software-based CFI scheme protecting the execution at basic block granularity of upcoming ARM-based devices against software \emph{and} fault attacks.
\fipac exploits ARM pointer authentication of recent ARMv8.6-A architectures to implement a cryptographically signed control-flow graph.
We cryptographically link the correct sequence of executed basic blocks to enforce control-flow integrity at this granularity at runtime.
We use a custom LLVM-based toolchain to automatically instrument programs without user interaction.
The evaluation on the application-grade SPEC2017 benchmark with different security policies shows a code size overhead between \speccoderange and a runtime overhead between \specruntimerange.
For small embedded benchmarks, we measured a runtime overhead between \embenchruntimerange.
While these overheads are significantly higher than for countermeasures against software attacks, \fipac outperforms related work protecting the control-flow against fault attacks.
\fipac is an efficient solution to provide protection against software- \emph{and} fault-based CFI attacks on basic block level on modern ARM devices.
\end{abstract}

\section{Introduction}


ARM-based systems are ubiquitous as billions of devices featuring such a processor are shipped, including mobile devices, the Internet-of-Things, or electronic control units in the automotive or industrial area.
This growing trend is continuing and ARM expects to embed up to a trillion cores over the next two decades~\cite{arm100bn}.
However, those devices are attacked using control-flow hijacks, posing a severe threat.
Those attacks bypass safety- and security-critical checks, such as privilege or password verifications and they also bypass additional countermeasures implemented in software.
Multiple attack methodologies covering different attacker models have been developed, which can induce control-flow hijacks.
While the control-flow can be manipulated entirely in software, \ie due to a memory vulnerability, also physical fault attacks can redirect the control-flow.
With the publication of software-induced fault attacks, the attack surface was increased even more, making many devices vulnerable to control-flow hijacks.

Physically or software-induced fault attacks can be used to manipulate the control-flow of a program.
For example, a recent work~\cite{rowhammerexploit} describes a NaCl sandbox exploit, where Rowhammer was used to manipulate the branch target of an indirect branch.
In~\cite{DBLP:conf/usenix/TatarKAGBR18}, a remote code execution exploit was crafted by inducing bitflips via the network interface on the global offset table of a program.
But also classical physical fault attacks are exploited to bypass security defenses.
There are several exploits where fault attacks are used to bypass secure boot~\cite{attackXbox360, DBLP:conf/fdtc/VasselleTMME17, DBLP:journals/tches/HerrewegenOGT21} or escalate Linux privileges and gain root access~\cite{DBLP:conf/ccs/VeenFLGMVBRG16, DBLP:conf/fdtc/TimmersM17}.

There also exist software-based attacks to manipulate the control-flow of a program.
Those attacks typically exploit a memory vulnerability to modify a code-pointer, to redirect the control-flow to a different location.
This strategy allows an adversary to perform powerful Turing-complete attacks, such as ROP~\cite{DBLP:conf/ccs/Shacham07} or JOP~\cite{DBLP:conf/ccs/CheckowayDDSSW10}.
These techniques have successfully been used to attack many different devices, from resource constraint embedded devices up to secure enclaves~\cite{DBLP:conf/ccs/JaloyanMARMN20,DBLP:conf/uss/HundHF09,DBLP:conf/uss/LeeJJKCCKPK17}.

To counteract fault and software-based threats on the control-flow on commodity devices, there exist different software-based countermeasures.
First, there are software CFI schemes~\cite{DBLP:conf/ccs/AbadiBEL05,DBLP:conf/osdi/KuznetsovSPCSS14,DBLP:conf/ccs/MashtizadehBBM15,DBLP:conf/uss/LiljestrandNWPE19}, which protect code-pointers and thereby enforce coarse-grained CFI.
Those coarse-grained protection schemes only protect indirect control-flow transfers, \ie indirect calls, but do not offer the needed protection against faults.
Second, there also exist more fine-grained countermeasures~\cite{DBLP:journals/tr/OhSM02a,DBLP:journals/tpds/AlkhalifaNKA99, DBLP:conf/dft/GoloubevaRRV03} to protect the control-flow of a program against fault attacks.
To cope with the performance penalty, these schemes protect the program execution at basic block granularity.
However, these protection mechanisms can easily be bypassed by a software attacker, as they are not considered in their threat model.

Although there exist countermeasures protecting against both threats, they require intrusive hardware changes~\cite{DBLP:conf/eurosp/WernerUSM18,DBLP:journals/compsec/ClercqGUMV17}.
Those schemes require to implement a custom processor, which is unrealistic for large-scale deployment, especially on closed architectures.
This leaves many applications exposed to software- or fault-based control-flow attacks.
Hence, it requires new and efficient countermeasures that automatically protect programs against both threats but without hardware changes.


\subsection*{Contribution}

In this work, we present \fipac, an efficient software-based CFI scheme protecting the execution at basic block granularity of upcoming ARM-based devices against software \emph{and} fault attacks.
\fipac's threat model considers a powerful attacker hijacking the control-flow on basic block level, independent of the used attack methodology.
We address this threat model and protect the control-flow by implementing a basic block level CFI protection scheme, which uses a keyed state update function resistant to memory vulnerabilities.

\fipac cryptographically links the sequence of basic blocks of a program at compile time.
During runtime, the countermeasure verifies that the executed sequence of basic blocks follows the statically determined control-flow graph~(CFG).
We exploit ARM pointer authentication of the ARMv8.6-A instruction set for efficient linking and verification.


We develop a custom toolchain based on LLVM to instrument and protect programs without user interaction against control-flow attacks on basic block level.
We validate the prototype implementation using a functional simulator supporting the ARMv8.6-A instruction set.
To evaluate the runtime performance of \fipac, we emulate the performance overhead of PA instructions and run SPEC2017 and other embedded benchmarks on existing hardware.
Moreover, we provide a security evaluation and discuss different security policies of \fipac.

Summarized, our contributions are:
\begin{itemize}
  \setlength\itemsep{0em}
  \item We present an efficient basic block granular CFI protection scheme for ARM-based systems protecting the control-flow against fault \emph{and} software attacks.
  \item We present a prototype implementation exploiting the ARM pointer authentication extension of the ARMv8.6-A instruction set.
  \item We provide a develop a custom LLVM-based toolchain to automatically instrument and protect arbitrary programs.
  \item We perform a functional and performance evaluation based on SPEC2017 and other embedded benchmarks and discuss different security policies.
\end{itemize}



\section{Background}
\label{sec:fipac:background}

In this section, we first give an introduction to fault attacks and different fault models and then discuss control-flow integrity as a countermeasure for software and fault attacks.

\subsection{Fault Attacks}

In a fault attack, the attacker influences the device's operating conditions with the goal of manipulating an inner state of the system.
Such a fault can be induced in different ways, \eg by manipulating the power supply~\cite{DBLP:conf/fc/BlomerS03,DBLP:conf/fdtc/BarenghiBPP09,DBLP:journals/tches/BozzatoFP19}, the clock signal~\cite{DBLP:conf/ches/PiretQ03,DBLP:journals/pieee/Bar-ElCNTW06}, the temperature~\cite{DBLP:conf/cardis/HutterS13}, or by shooting with a laser or electromagnetic impulse onto the chip surface~\cite{DBLP:conf/crypto/BihamS97,DBLP:conf/ches/SkorobogatovA02,DBLP:conf/fdtc/MoroDHRE13,DBLP:conf/cardis/SelmkeBHS15}.
While these fault methodologies require physical access to the device, more recent attacks have shown that this constraint can be relaxed.
For example, new methodologies, such as the Rowhammer effect~\cite{DBLP:conf/isca/KimDKFLLWLM14}, can be used to manipulate bits in memory by frequently accessing neighboring memory cells.
This behavior can even be exploited remotely via Javascript~\cite{DBLP:conf/dimva/GrussMM16} or over the network interface~\cite{DBLP:journals/corr/abs-1805-04956,DBLP:conf/usenix/TatarKAGBR18}.
The growing number of new fault attacks, such as Plundervolt~\cite{DBLP:conf/sp/MurdockOGBGP20}, VoltJockey~\cite{DBLP:conf/ccs/QiuWLQ19, DBLP:conf/asianhost/QiuWLQ19, DBLP:journals/sigmobile/QuiWLQ20}, or CLKscrew~\cite{DBLP:conf/uss/TangSS17}, highlight the severity of software-induced fault attacks on commodity devices.
Irrespective of the used fault injection methodology, the fault model defines if the fault either targets to tamper data or the control-flow of a program.

\paragraph{Data Attacks.}

Historically, fault attacks were mainly used to break cryptographic primitives by injecting a fault into the computation of the target~\cite{DBLP:conf/eurocrypt/BonehDL97, DBLP:conf/crypto/BihamS97, DBLP:conf/wistp/TunstallMA11, DBLP:journals/tches/DobraunigEKMMP18}.
In general, in these attacks, the injected fault aims to alter data, the internal cryptographic state, to weaken security guarantees of the cryptographic scheme.
For example, such data attacks allow an adversary to extract AES keys or break RSA authentication~\cite{DBLP:conf/ccs/QiuWLQ19, DBLP:conf/asianhost/QiuWLQ19, DBLP:journals/sigmobile/QuiWLQ20, DBLP:conf/uss/TangSS17} using software-induced faults on commodity devices featuring an ARM processor.
To counteract these attacks, data redundancy schemes~\cite{hamming1950error, DBLP:conf/ches/JoshiWK04, DBLP:journals/tc/BertoniBKMP03, DBLP:conf/dft/AnsariV10, langdon1970concurrent} have been developed, which are capable of detecting such faults on data.

\paragraph{Control-Flow Attacks.}

Although data protection schemes can thwart attacks on a large class of fault attack targets, they cannot prevent an attacker from hijacking the control-flow of a program using a fault.
In this fault model, the adversary aims to bypass security critical checks, such as secure boot or other authentication mechanisms, or unconditionally jump to sensitive code blocks.
These attacks, which are typically conducted by corrupting instructions, target to manipulate the program within (intra) or over (inter) a basic block, \ie consecutive instruction sequence without a control-flow transfer.
While intra basic block attacks allow the attacker to skip or manipulate individual instructions within a basic block, inter basic block attacks enable the attacker to redirect the control-flow to an arbitrary code position by corrupting addresses of calls.

\subsection{Control-Flow Integrity}
\label{sec:fipac:cfi}

To protect a program from intra or inter basic block control-flow attacks, enforcing control-flow integrity has shown to be an effective defense~\cite{DBLP:journals/tissec/AbadiBEL09}.
However, existing software-based CFI protection schemes are implemented with different enforcement granularities and typically either address a software \emph{or} a fault attacker but not both.
Although there exist CFI schemes addressing both threats, these schemes require intrusive hardware changes, also leading to significant performance penalties, which are not feasible for off-the-shelf hardware.

\subsubsection{Software CFI Schemes}
\label{sec:fipac:scfi}

Software CFI~(SCFI)~\cite{DBLP:journals/tissec/AbadiBEL09} protects the program from a software adversary performing control-flow hijacking attacks.
This countermeasure enforces the control-flow graph~(CFG) extracted at compile time by dynamically protecting a subset of inter basic block control-flow transfers at runtime.
The coarse-grained CFI policy, which only protects indirect function calls or returns, can either be inserted during the compilation~\cite{DBLP:conf/uss/TiceRCCELP14} or it can be applied to compiled binaries~\cite{DBLP:conf/uss/ZhangS13}.
To improve the performance of such countermeasures and make the deployment more practical, the CFI policy was gradually relaxed.
CPI~\cite{DBLP:conf/sp/EvansFGOTSSRO15} and CCFI~\cite{DBLP:conf/ccs/MashtizadehBBM15} protect a broad range of forward- and backward-edges of the program by maintaining the integrity of all code-pointers in the program.
Similarly, PARTS~\cite{DBLP:conf/uss/LiljestrandNWPE19} protects code-pointers by signing and verifying the code-pointers using ARM's pointer authentication feature before using them.
If the verification fails, \ie the pointer authentication code stored in the pointer does not match the expected pointer authentication code, the application is stopped.
PACStack~\cite{DBLP:journals/corr/abs-1905-10242}, a more coarse-grained CFI policy, protects return addresses on the stack by utilizing the PA feature to cryptographically link and verify those addresses.

\subsubsection{Fault CFI Schemes}
\label{sec:fipac:fcfi}

Contrary to SCFI, fault CFI schemes~(FCFI) consider a hardware attacker performing fault attacks in their threat model and, therefore, operate on a much finer granularity.
FCFI schemes capable of detecting intra basic block control-flow hijacks, \eg instruction skips, employ a global CFI state, which uniquely gets updated with the execution of each instruction.
However, as maintaining and checking a state at this granularity is expensive, these schemes require dedicated hardware changes for performance reasons~\cite{DBLP:conf/eurosp/WernerUSM18,DBLP:conf/cardis/WernerWM15,DBLP:journals/compsec/ClercqGUMV17,DBLP:journals/ieicet/Sugihara11}.
As intrusive hardware changes are not possible for commodity devices, software-based FCFI schemes provide an efficient trade-off between security and performance by protecting all control-flow transitions between basic blocks, hence, providing full inter basic block CFI protection.
In CFCSS~\cite{DBLP:journals/tr/OhSM02a} and SWIFT~\cite{DBLP:conf/cgo/ReisCVRA05}, each basic block is assigned a unique signature at compile-time.
When entering the basic block, the global control-flow state is updated with this signature and is compared to match the expected state at this location.

\begin{algorithm}
  \caption{CFI state update function.}
  \label{alg:fipac:state_update}
  \begin{algorithmic}[1]
    \Function{Update}{$S$, Sig$_{BB}$}
      \State $r_1$ $\gets$ Sig$_{BB}$
      \State $S$ \,$\gets$ $S$ $\oplus$ $r_1$
    \EndFunction
  \end{algorithmic}
\end{algorithm}
This control-flow state update, as shown in Pseudocode~\ref{alg:fipac:state_update}, can be implemented with an XOR operation, like in CFCSS~\cite{DBLP:journals/tr/OhSM02a}.
Here, the global CFI state $S$ is XORed with the signature Sig$_{BB}$ of the corresponding basic block, which is stored in the binary.
At certain locations of the programs, check operations are included, which compare this CFI state to the expected value.
This approach of CFCSS yields a runtime overhead between 107\,\% and 426\,\%~\cite{DBLP:conf/dft/GoloubevaRRV03}.
ACFC~\cite{DBLP:conf/iolts/VenkatasubramanianHM03} reduces the performance penalty down to 47\,\% by decreasing the checking precision and thereby reducing the security guarantees.
Similarly, other approaches~\cite{DBLP:conf/esorics/LalandeHB14, DBLP:journals/compsec/HeydemannLB19} annotate the source code with counter increment and verification macros to detect control-flow deviations.
However, a protection scheme requiring manual source code modifications is not practical.
It cannot easily protect legacy code and it makes a large-scale deployment hard.

\section{Threat Model and Attack Scenario}
\label{sec:fipac:threat}

In this section, we first present the threat model of this work and show how this threat model can bypass existing SCFI and FCFI protection schemes.
Finally, we discuss the required properties for a secure CFI scheme protecting against software and fault attackers.

\subsection{Threat Model}

\fipac considers an attacker capable of performing software and fault attacks to redirect the control-flow of a program.
This attacker aims to hijack direct or indirect control-flow transfers, \ie the threat model of \fipac covers all transfers between basic blocks of the program.
We consider attacks on the control-flow independently of the used methodology, \ie we cover physical or software-induced fault attacks or classic software attacks.
We assume the attacker has full read access to the binary and can read all instructions and data.
This threat model includes a software attacker, which uses this information to exploit a memory vulnerability with the goal of conducting a control-flow hijack, \eg manipulate code-pointers to perform a ROP or JOP exploit.
Furthermore, we assume ARM pointer authentication to be cryptographically secure and that its keys are isolated from user applications.

As we only consider control-flow hijacks on the edges of the CFG, we exclude attacks within a basic block, \eg instruction skips.
However, we highlight in Section~\ref{sec:fipac:futurew}, how \fipac also could protect sensitive code pieces at instruction granularity using manual code instrumentation.
Furthermore, fault attacks on the data or on the actual computation are not in the scope of this work, which includes data used during a conditional branch.
It requires orthogonal countermeasures, \eg a data encoding scheme or instruction replication, to protect the actually processed data and its computation.
For a full protection against fault attacks, a combination of both, the protection of data and processing and a control-flow protection such as \fipac is required.
We now show how existing CFI protection schemes can be bypassed in the stated threat model.




\paragraph{Fault Attacks on SCFI.}

Most software CFI schemes~\cite{DBLP:journals/tissec/AbadiBEL09, DBLP:conf/uss/TiceRCCELP14, DBLP:conf/uss/ZhangS13, DBLP:conf/sp/EvansFGOTSSRO15, DBLP:conf/uss/LiljestrandNWPE19, DBLP:journals/corr/abs-1905-10242} do not consider a fault attacker in their threat model and, therefore, can be bypassed with a single, targeted fault.
As the code section of the program is assumed to be immutable and cannot be altered by an adversary, SCFI schemes only protect indirect control-flow transfers but not direct calls and other branches.
Hence, a targeted fault to the code segment of a program or directly within the execution, \eg a fault on the program counter or on the immediate value of a direct call, cannot be detected by SCFI.

\paragraph{Bypassing FCFI.}

The threat models of software-based FCFI schemes do not consider classical software attackers.
Contrary to software CFI threat models~\cite{DBLP:journals/tissec/AbadiBEL09}, where the memory is considered to be vulnerable, typical fault CFI schemes do not include this scenario in their threat model.
An attacker exploiting a memory vulnerability can easily tamper the CFI state, which is maintained purely in software.
Since the CFI state update function is known to the adversary, an attacker-controlled global CFI state can be crafted.
Even a naïve combination of SCFI and FCFI, which are secure in their own threat model, can be bypassed, as discussed in Section~\ref{sec:fipac:secomparison}, with a combined fault and software attack.

\begin{figure*}[t]
  \centering
  \begin{subfigure}[b]{0.32\textwidth}
      \centering
      \includegraphics[width=\textwidth]{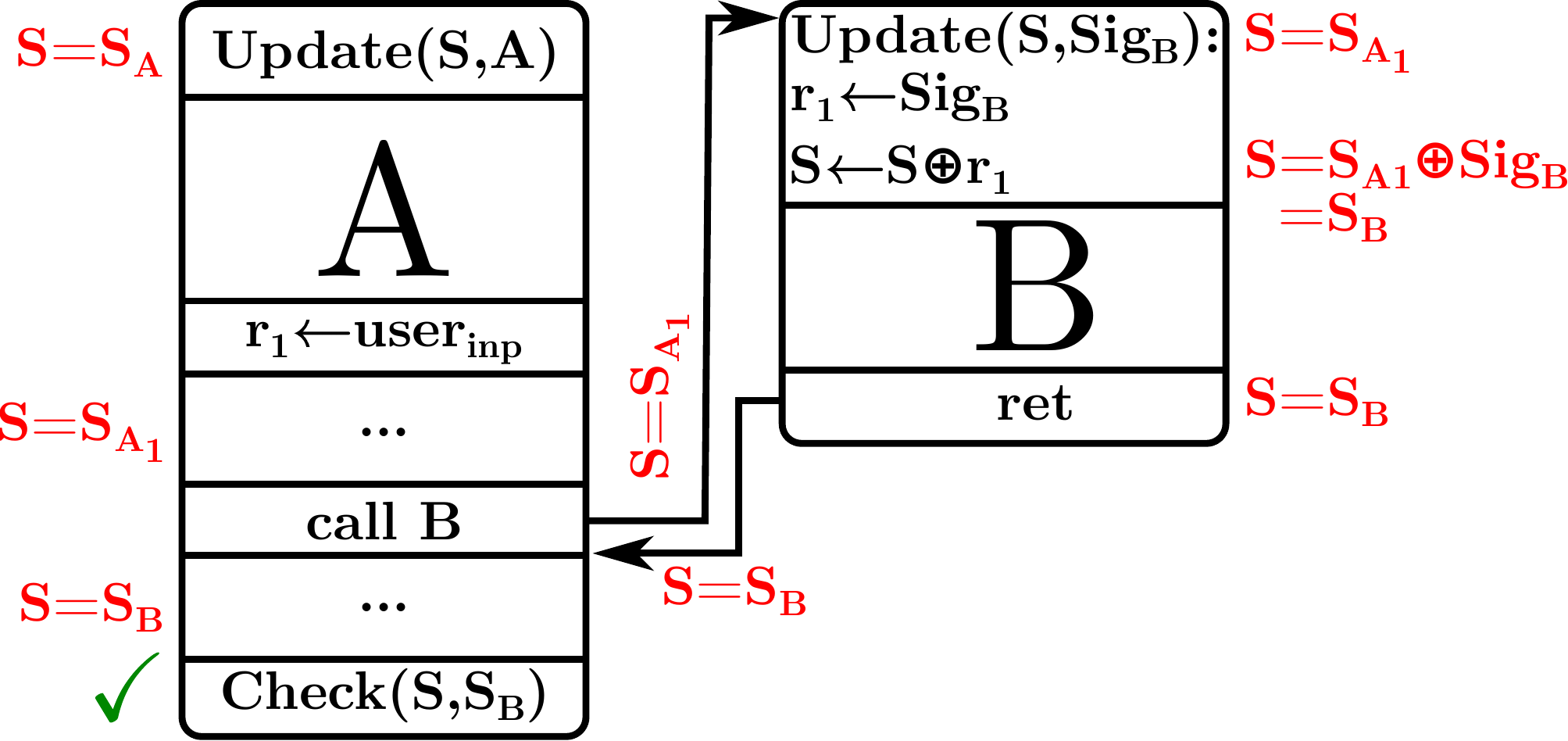}
      \caption{Valid control-flow.}
      \label{fig:fipac:noattack}
  \end{subfigure}
  \hfill
  \begin{subfigure}[b]{0.32\textwidth}
      \centering
      \includegraphics[width=\textwidth]{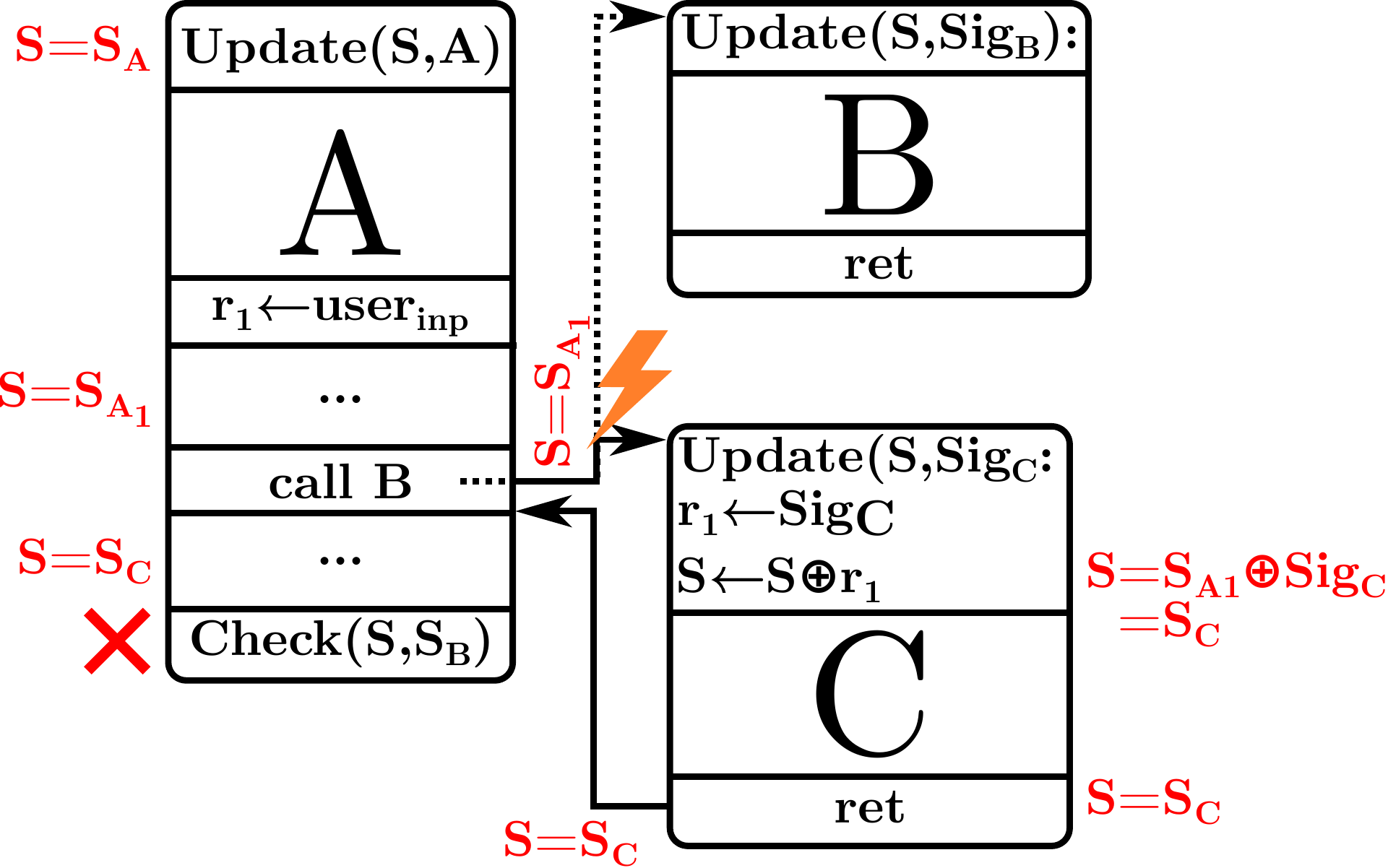}
      \caption{Detectable attack.}
      \label{fig:fipac:simpleattack}
  \end{subfigure}
  \hfill
  \begin{subfigure}[b]{0.32\textwidth}
      \centering
      \includegraphics[width=\textwidth]{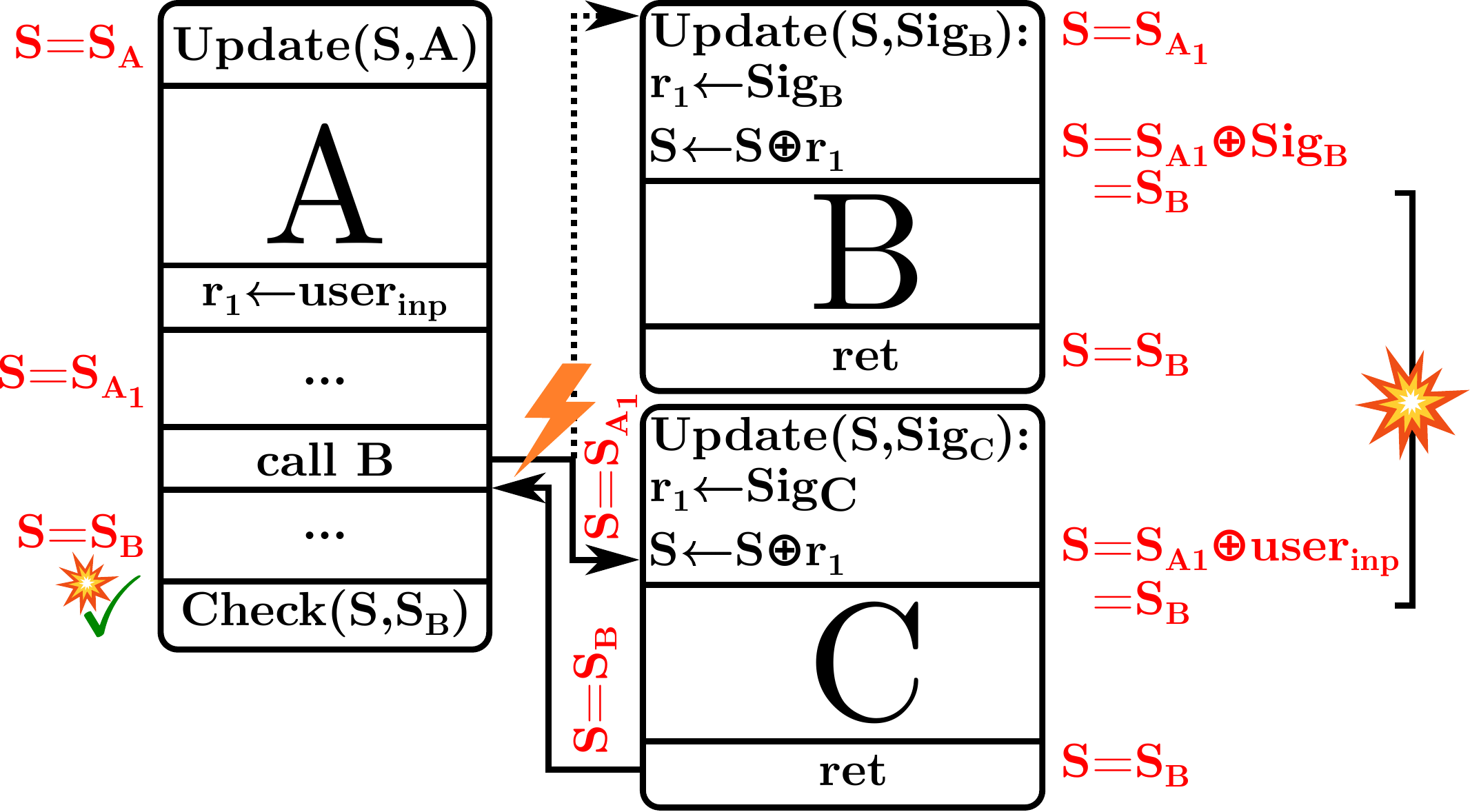}
      \caption{Successful attack.}
      \label{fig:fipac:combinedattack}
  \end{subfigure}
  \caption{Attacker scenario to bypass FCFI.}
  \label{fig:fipac:attackscenario}
\end{figure*}

To highlight the conceptional weaknesses of FCFI schemes, we demonstrate a possible attack bypassing FCFI with its state update function as shown in Pseudocode~\ref{alg:fipac:state_update}.
Such a state update function is similar for many software-based CFI protection schemes~\cite{DBLP:journals/tr/OhSM02a}.
They all compute their CFI signatures in software and load them into a register at some point in time.
The goal is to exploit this instruction sequence of the state update function to manipulate the global CFI state to an attacker defined value, \ie bypassing CFI.

Figure~\ref{fig:fipac:attackscenario} depicts the attacker scenario for a control-flow hijacking attack.
Without an attacker present, Figure~\ref{fig:fipac:noattack} shows a valid control-flow transfer, where \texttt{A} calls \texttt{B}.
When entering \texttt{B}, the state update function updates the global state $S$ to the beginning state $S_{B}$ by XORing the signature $Sig_{B}$ to the previous state $S$.
After returning from \texttt{B}, a CFI check verifies the state $S$ to be the pre-computed state $S_B$.

In Figure~\ref{fig:fipac:simpleattack}, we now consider an attacker redirecting the control-flow of the call from \texttt{B} to \texttt{C}.
At the beginning of \texttt{C}, the state update function again XORs the current state $S$ with the signature of function \texttt{C}.
However, as this state $S=S_C$ deviates from the pre-computed state $S_B$, the control-flow hijack can be detected in the final check.

In the last scenario, Figure~\ref{fig:fipac:combinedattack} shows a successful attack on the control-flow, bypassing FCFI.
Here, register $r_1$ is controlled by the attacker, \eg it is used to store user input, it is modified due to a memory vulnerability, or it is manipulated with a targeted fault.
The adversary again redirects the control-flow from \texttt{B} to \texttt{C}, but omits the signature load to $r_1$.
Since $r_1$ is controlled by the attacker who knows all states and signatures, the final state of function \texttt{C} can be forged to match the end state of function \texttt{B}.
Eventually, the final CFI check in function \texttt{A} cannot detect the control-flow hijack.
Note, the control-flow redirect in Figure~\ref{fig:fipac:simpleattack} or \ref{fig:fipac:combinedattack} can either be performed with a software attack or by inducing a fault.

\subsection{Requirements for CFI against Software and Fault Attacks}
\label{sec:fipac:sfcfi}

To protect against both, software and fault attacks, and to enable a large-scale deployment, CFI protection schemes need to fulfill the following four requirements:
\begin{enumerate}
  \setlength\itemsep{0em}
  \item The CFI protection needs to enforce the CFI at a fine granularity, \ie at least on basic block level, to protect from a fault attacker.
  \item The proper selection of the CFI state update function is essential, as it directly influences the security of the CFI scheme.
  Choosing a weak state update function, \eg a simple XOR, allows an attacker to bypass the control-flow protection.
  An attacker reading the binary is able to recompute the CFI states for all locations of the program and can bypass the CFI scheme, as discussed in Section~\ref{sec:fipac:fcfi}.
  Thus, it is required that CFI states are not known to the attacker and cannot be recomputed even when having full access to the binary.
  Furthermore, the state update function must be accumulating, meaning that the next CFI state depends on the value of the previous CFI state.
  \item To make the protection deployable for a wide range of devices, the protection should not require hardware changes and can be implemented in software.
  \item To support legacy code bases and to enable an easy deployment, the CFI protection must be applied automatically, \ie during compilation, and must not require source code modifications
\end{enumerate}
Previous fine-grained CFI protection schemes with keyed update functions require expensive hardware changes and are not suitable for commodity devices.
In SCFP~\cite{DBLP:conf/eurosp/WernerUSM18} and in SOFIA~\cite{DBLP:journals/compsec/ClercqGUMV17}, the program is encrypted at compile time and the instructions are decrypted at runtime using control-flow dependent state information.
However, both schemes require intrusive hardware changes in the instruction pipeline of the processor and are therefore inapplicable for large-scale deployment.
Hence, there is a need for efficient CFI protection schemes considering software and fault attacks, which do not require hardware changes.


\section{\fipac}
\label{sec:fipac:design}

In this section, we present \fipac, an efficient software-based CFI solution for ARM-based devices, fulfilling the key requirements discussed above.
We first show the state-based CFI concept based on the work of Wilken and Shen~\cite{DBLP:conf/itc/WilkenS88, DBLP:journals/tcad/WilkenS90} and then discuss how more advanced control-flow operations are protected.
Finally, we discuss the selection of the CFI state update function and the placement of CFI checks within the program.

\subsection{Signature-Based Control-Flow Integrity}

\fipac is a state-based CFI protection scheme, where every location in the program corresponds to a well-defined CFI state.
This state is maintained globally through the execution of a program.
At certain locations in the program, the CFI state is checked to match the expected execution state, indicating that no control-flow error occurred.
To consider the history of the execution-flow, the next CFI state is linked with the previous one, which allows \fipac to enforce the CFG of a program.

Typical programs do not have a simple, linear control-flow, but contain control-flow transfers, such as conditional branches, loops, or function calls.
Depending on which path of the program is executed, the CFI state for a certain basic block would differ since it has more than one predecessor.
When the control-flow merges, \ie for conditional branches, two different paths of CFI states merge and would turn into a state collision.
To avoid these problems, we adapt the concept of generalized path signature analysis from  Wilken and Shen~\cite{DBLP:conf/itc/WilkenS88, DBLP:journals/tcad/WilkenS90} and insert so-called justifying signatures to correct the state.

\begin{figure*}[t]
  \begin{minipage}[b]{0.4\textwidth}
      \centering
        \includegraphics[width=\textwidth]{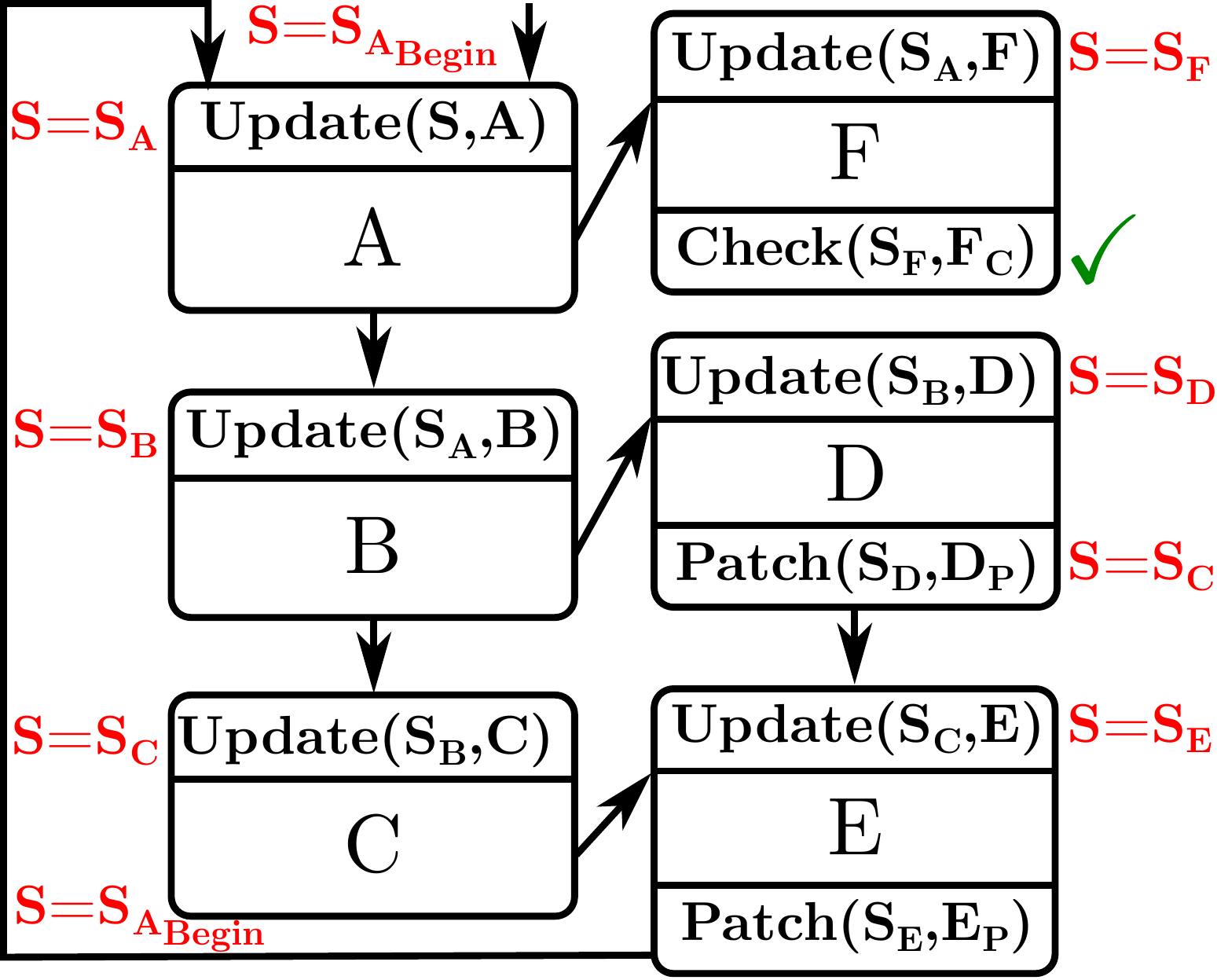} \caption{Justifying signature to support control-flow merges.}
        \label{fig:fipac:patch_conditional_branch}
  \end{minipage}
  \hfill
  \begin{minipage}[b]{0.5\textwidth}
      \centering
      \includegraphics[width=\textwidth]{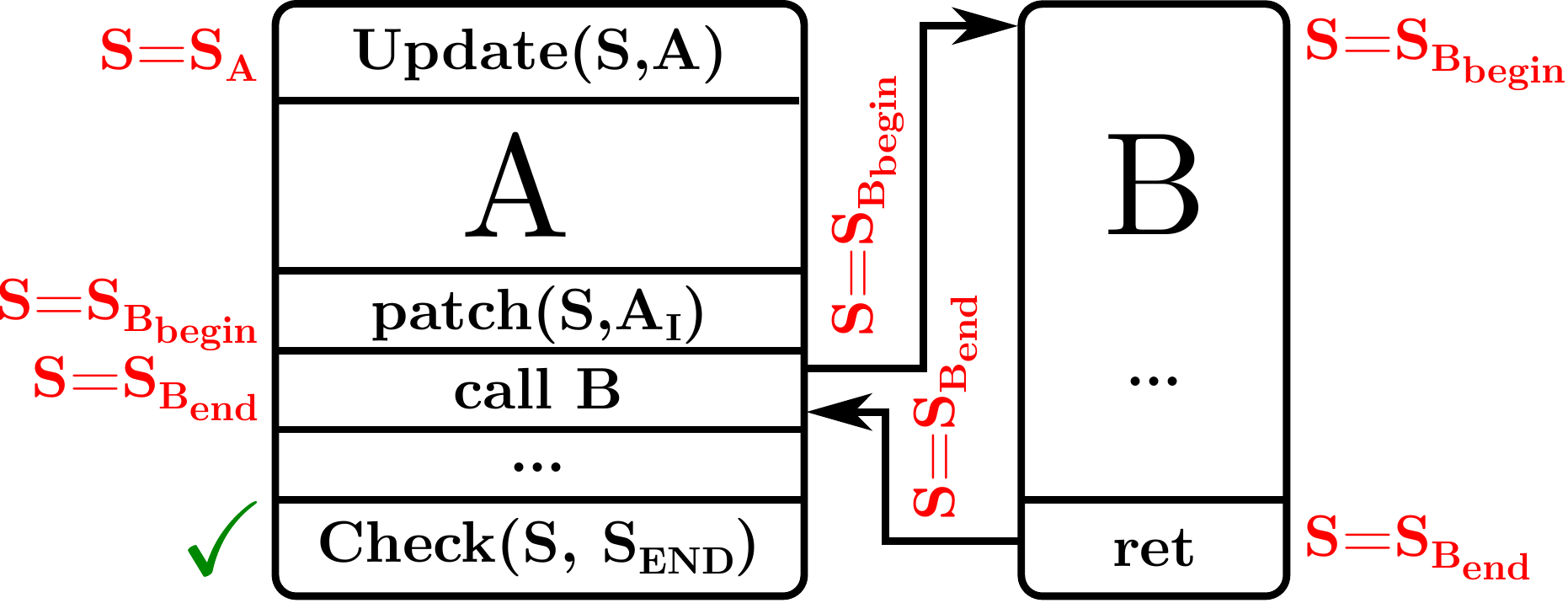} \caption{CFI state patch for direct calls.}
      \label{fig:fipac:patch_direct_call}
  \end{minipage}
\end{figure*}

Control-flow merges occur when the program contains conditional branches, diverting the control flow into two paths.
In Figure~\ref{fig:fipac:patch_conditional_branch}, we show a conditional branch, where the control-flow merges in basic block \texttt{E} and a loop, which control-flow merges in basic block \texttt{A}.
At the end of the basic block \texttt{D}, there is a single state update that ensures that the CFI state at the beginning of basic block \texttt{E} is the same, whether coming from basic block \texttt{C} or \texttt{D}.
These update values are stored in the program code and are applied at runtime.
In Figure~\ref{fig:fipac:patch_conditional_branch}, basic block \texttt{E} jumps back to \texttt{A}, thus forming a loop.
For this reason, a patch update is inserted at the end of basic block \texttt{E}, which corrects the CFI state to $S_{A_{Begin}}$.

While hardware-centric approaches typically update the CFI state on instruction granularity, this approach is too expensive in software.
Thus, \fipac, our software-centric design, reduces the CFI state update granularity to basic block level, which is reasonable in terms of performance but still provides full protection of the CFG on basic block level.

\subsection{Direct Calls}


Direct function calls require special handling of the justifying signatures.
As shown in Figure~\ref{fig:fipac:patch_direct_call}, function \texttt{A} directly calls function \texttt{B}.
To support calling function \texttt{B} from multiple call sites, the beginning CFI state of function \texttt{B} always needs to be the same.
For this reason, we apply a justifying signature at the call site before the direct call, which transforms the call site's CFI state to the beginning state of function \texttt{B}.
When returning, the CFI state continues with the end state of the called function, in this example $S_{B_{end}}$.

\subsection{Indirect Calls}
\label{sec:fipac:design:icalls}

\begin{figure*}[t]
  \begin{center}
    \setlength{\belowcaptionskip}{-5mm}
    \includegraphics[width=0.95\linewidth]{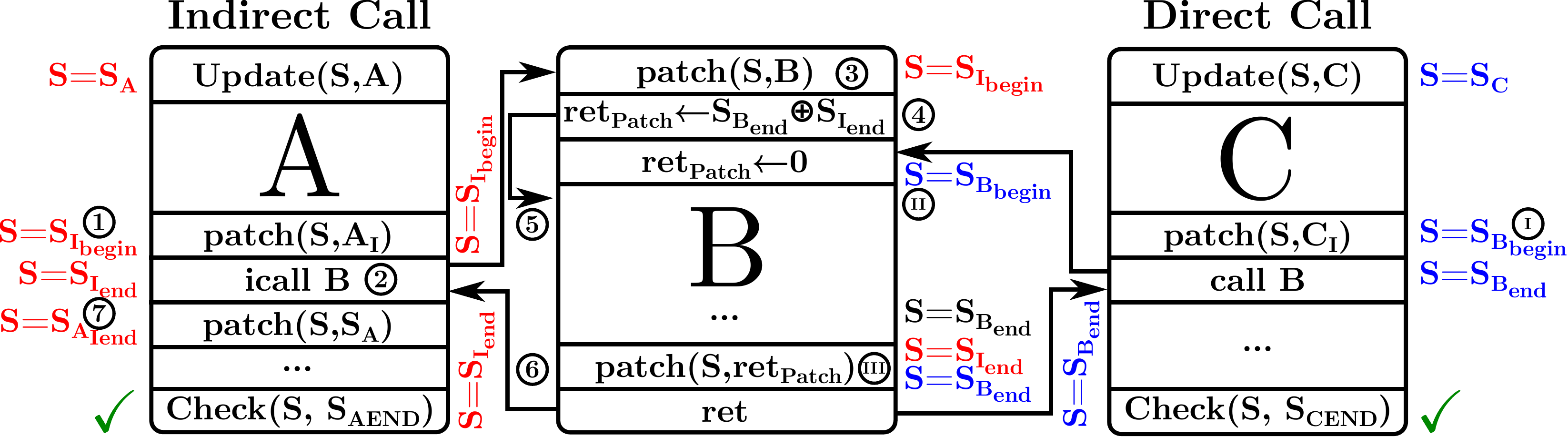}
    \caption{CFI state patch for indirect calls.}
    \label{fig:fipac:patch_indirect_call}
  \end{center}
\end{figure*}

Indirect function calls require special handling of justifying signatures and are not covered by the work of Wilken and Shen.
Determining the exact function that is being called during the indirect call is not always possible at compile time.
Indirect function calls can also call different functions from the same call site, \eg a function pointer is given as a parameter to another function and is then called inside.
The best that \fipac can do to protect indirect calls is determining a possibly over-approximated set of potential call targets.
Then, \fipac enforces that the indirect call can only call one of these functions.
Figure~\ref{fig:fipac:patch_indirect_call} shows the patching for indirect calls and how this interacts with direct calls.

To support the CFI protection of indirect calls, \fipac determines an intermediate CFI state $S_I$ for every set of indirectly called functions.
Note, this can also lead to merging sets if the same function is called indirectly from different call sites.
When performing an indirect call, the call site \texttt{A}, in \circleds{1}, first patches its CFI state $S_A$ to an intermediate state $S_{I_{begin}}$, which is the same for all possible call targets of this indirect call.
Next, in \circleds{2}, the indirect call is performed.
At the beginning of the indirectly called function \texttt{B}, we transform the CFI state, in \circleds{3}, from the common intermediate state $S_{I_{begin}}$ to the beginning state of function $S_{B_{begin}}$.
Furthermore, in \circleds{4}, we set up the patch value used for the function return.
We jump over the direct call function header in \circleds{5} and then continue the execution of function \texttt{B} until the return patch in \circleds{6}.
This patch transforms the CFI end state $S_{B_{End}}$ of function \texttt{B} to the common intermediate return state $S_{I_{End}}$ followed by a return to the caller.
The caller \texttt{A} then uses the pre-call signature $S_A$, which was saved on the stack, for a state update in \circleds{7}, to transform the intermediate return state to a unique continuing state for \texttt{A}.
Note, the call site could simply continue with the execution using the CFI state $S_{I_{end}}$.
However, this would introduce undetectable control-flow vulnerabilities between different indirect call sites of the same function.
Therefore, the patch with $S_A$ is necessary, to avoid different call sites to continue with the same signature, as well as to ensure that the function was actually called.
The call site continues with the execution using the CFI state $S_{A_{Iend}} = S_{I_{end}} \oplus S_A$, which is different for every call site.

Since any function must be callable with direct or indirect calls, the handling of indirect and direct calls interacts.
On the right of Figure~\ref{fig:fipac:patch_indirect_call}, we show how \texttt{C} calls function \texttt{B} directly.
In \circleds{I}, a justifying signature is applied to transform \texttt{C}'s CFI state to the beginning state $S_{B_{begin}}$ of function \texttt{B}.
The direct call does not jump to the beginning of function \texttt{B}, instead, it jumps to a dedicated entrypoint setting up the return patch $ret_{Patch}$ to be zero (\circleds{II}).
Then, it normally continues with the execution of function \texttt{B}.
At the end of the function in \circleds{III}, the return $ret_{Patch}$ is applied to the state.
Since the return patch value is zero, this statement does not have an effect on the CFI state, which remains at $S_{B_{end}}$.
After the return, the call site then continues with the execution using the CFI state $S_{B_{end}}$.

\subsection{State Updates using ARM Pointer Authentication}

As discussed in Section~\ref{sec:fipac:sfcfi}, the state must not be computable by the attacker and must depend on all previous CFI states.
To solve this problem, \fipac uses a chained cryptographic message authentication code~(MAC) for the CFI state update function.
Thereby, we bind the security of the CFI scheme to a secret cryptographic key, which is unknown to the attacker and isolated by the operating system.

To efficiently implement such a cryptographic function, we exploit ARM pointer authentication~(PA), which was introduced in ARMv8.3-A and updated it in ARMv8.6-A with \textit{EnhancedPAC2} and \textit{FPAC}~\cite{armv8}.
This feature is designed to cryptographically sign pointers with a pointer authentication code~(PAC) and verify its integrity before using it~\cite{armpac}.
Internally, the pointer authentication code is generated by computing the MAC over the pointer and a modifier register.
The MAC is computed using the QARMA~\cite{DBLP:journals/tosc/Avanzi17} block cipher with the modifier register being an additional input to the tweakable block cipher.
The combination of modifier and five different keys allow the protection of different domains.
Although pointers are 64-bit values, the size of the virtual address space limits the actual size of the pointer values.
In AArch64 Linux, the virtual address space is typically configured for 39 or 48-bit~\cite{aarch64va}, leaving the upper bits pointers unused.
ARM PA uses these unused bits to efficiently store the PAC value in the upper unused bit, thus having no additional storage overhead.

To use pointer authentication, the ARMv8.6-A the instruction set was extended for computing and verifying a PAC.
With EnhancedPAC2, the instructions \texttt{PACI*} and \texttt{PACD*} with a given key \texttt{A} or \texttt{B} use the destination register as input address, the source register as a modifier, and XOR the PAC in the upper bits of the destination register.
The PAC in the pointer can be verified by using the \texttt{AUTI*} and \texttt{AUTD*} instructions.
On a successful verification, the PAC is removed from the address, and the pointer can be used.
If the verification fails, the \texttt{AUT*} instructions with FPAC generate a trap (this is different compared to ARMv8.3-A, which only set an error bit in the PAC).




In this work, we use the PA mechanism of ARMv8.6-A to implement the state update function rather than sign pointers.
This instruction set extension fulfills the requirements needed for the state update function.
It uses a keyed mechanism, and with ARM EnhancedPAC2, it also brings in the accumulating functionality required for linking subsequent states.
We further discuss the use of PA instructions in Section~\ref{sec:fipac:sysimplementation}.

\subsection{Placement of Checks}

Although the placement of CFI checks is essential for the security of the CFI scheme, there is no general solution for the correct placement of checks.
However, at minimum, there needs to be one check at the end of the program.
For programs that do not return, \ie server programs, at least one CFI check in the main event loop is needed.
This strategy, however, has the longest detection latency and the worst detection probability.
To reduce the detection latency and improve the detection probability of CFI errors, more CFI checks are required.
However, the granularity is always a trade-off between runtime overhead and security.
The more checks inserted, the more overhead but also better detection probability and lower detection latency.
At worst, a CFI check is placed at the end of every basic block, yielding the best security but worst runtime and code performance.

In between, there exist arbitrary policies of placing a CFI check with different trade-offs.
For example, a generic intermediate policy places a CFI check at the end of each function.
Note, also fully custom strategies of placing checks are possible.
With the help of dynamic runtime profiling of an application, an optimized compiler can place the checks more efficiently.
For example, a check policy can place a CFI check after every 100\textsuperscript{th} basic block.

\section{Implementation}
\label{sec:fipac:implementation}

In this section, we discuss the prototype implementation of \fipac based on the ARMv8.6-A architecture and discuss the custom LLVM-based toolchain.

\subsection{System Implementation}
\label{sec:fipac:sysimplementation}

\fipac computes a rolling CFI signature throughout the execution of the program.
We implement \fipac in software on top of the ARMv8.6-A architecture in the AArch64 execution state without any hardware changes.
\fipac exploits the updated PA instruction set extension to implement the cryptographic state update function.
The \texttt{PACI*} and \texttt{PACD*} instructions cryptographically compute a MAC over a given pointer and a modifier register and store the result in the upper bits of the pointer.
In ARMv8.6-A with \textit{EnhancedPAC2}, these instructions do not simply replace the upper bits of the pointer with the computed MAC but instead XOR them to the existing upper bits, as shown in Pseudocode~\ref{alg:fipac:pacia}.

\begin{algorithm}
  \caption{Behavior of the \pacia instruction in the 16-bit configuration.}
  \label{alg:fipac:pacia}
  \begin{algorithmic}[1]
    \Function{pacia}{Xd, Xm}
      \State PAC[63:0] $\gets$ ComputePAC(Xd[47:0], Xm, K)
      \State Xd[63:48] $\gets$ Xd[63:48] $\oplus$ PAC[63:48]
      \State Xd[47:0]  $\text{ }\gets$ Xd[47:0]
    \EndFunction
  \end{algorithmic}
\end{algorithm}

Instead of signing a pointer with \pacia, we use this instruction in \fipac to compute the CFI state.
Here, the upper bits of a \pacia computation, the PAC bits, denote our CFI state.
The size of the PAC depends on the virtual memory configuration.
In this instantiation of \fipac, we use a 16-bit configuration for the PAC.
To accumulate the CFI state, the \pacia instruction is always executed on the same \enquote{pointer}, in our case, the CFI state stored in \texttt{Xd}.
For each basic block, a unique identifier, \ie the program counter~(PC), is used as the modifier \texttt{Xm} for this instruction.
By subsequently XORing the new CFI state to the previous one, we create a dependency link between succeeding basic blocks.
We store global CFI state in the exclusively reserved the general purpose register \texttt{x28}.


\paragraph{State Updates and Patches.}

\begin{figure*}[t!]
  \begin{minipage}[t]{.28\textwidth}
    \begin{lstlisting}[language=c++,style=aarch64,numbers=none,caption={CFI state update with \pacia.},label={lst:fipac:pac_state_update},belowskip=-6mm]

adr   x2,  #4
pacia x28, x2 \end{lstlisting}
  \end{minipage}
  \begin{minipage}[t]{.28\textwidth}
    \begin{lstlisting}[language=c++,style=aarch64,numbers=none,caption={CFI state patch.},label={lst:fipac:pac_state_patch},belowskip=-6mm]

mov  x2, #patch
eor x28, x28, x2 \end{lstlisting}
  \end{minipage}
  \begin{minipage}[t]{.28\textwidth}
    \begin{lstlisting}[language=c++,style=aarch64,numbers=none,caption={CFI check with \autiza.},label={lst:fipac:pac_check},belowskip=-6mm]
mov    x2, #const
eor    x2, x28, x2
autiza x2  \end{lstlisting}
  \end{minipage}
\end{figure*}

To cryptographically compute the CFI state, we exploit the updated PA extension of ARMv8.6-A.
The \texttt{PACIA, Xd, Xn} instruction computes a PAC of register \texttt{Xd} with \texttt{Xn} as the modifier and XORs it to the upper bits of \texttt{Xd}.
In Listing~\ref{lst:fipac:pac_state_update}, we show the CFI state update, which is placed at the beginning of each basic block.
By using \texttt{ADR, x2, \#4}, we first load a unique constant for the basic block to a temporary register \texttt{x2}, in this case the address of the instruction.
Second, this constant is used to compute a new, valid PAC, which gets XORed to the previous CFI state in \texttt{x28}.

As discussed before, it requires justifying signatures for control-flow transfers such as conditional branches or calls.
In Listing~\ref{lst:fipac:pac_state_patch}, we show the instructions for such a patch update.
We load an immediate constant to a temporary register in \texttt{x2}, which gets XORed to the CFI state in \texttt{x28}, thus correcting it to a target state.
Note that an XOR applied to a valid PAC value generates an invalid PAC, which needs to be corrected before verification.

\paragraph{Checks.}

A CFI check performs a comparison of the current CFI state with the expected state at this location and executes an error handler on a mismatch.
However, such instruction sequences typically involve conditional branches, which slows down the execution of a program, as they have a negative impact on the instruction pipeline.
Hence, we also exploit the PA instruction set extension for efficiently performing the necessary CFI checks.

Similar to generating a PAC value, the ARM also provides \texttt{AUTI*} and \texttt{AUTD*} instructions to verify the integrity of PAC values.
In ARMv8.6-A with \textit{FPAC}, these instructions trap on an invalid PAC verification.
Since we use \pacia to compute a PAC, it is tempting to directly use the corresponding \autiza instruction for verification.
However, the running CFI state in \texttt{x28} is not a valid PAC value in the classical sense.
Instead, the CFI state is an accumulated XOR-sum of many valid PAC values, which combined together do not form a valid PAC anymore.
Therefore, we cannot directly use the \autiza instruction to check the control-flow integrity.

However, at every location in the program, we know the expected CFI state at compile time.
Thus, we can also compute a constant, which XORs the CFI state at the location of the check to a valid PAC.
This constant is determined in the post-processing tool and explained in Section~\ref{sec:fipac:pp}.
By applying this constant to the global CFI state, we receive a valid PAC that can be verified with the \autiza instruction.
Although the \autiza instruction does not use a modifier, in comparison to \pacia, we choose the constant correction value in such a way, that a verifiable PAC is generated.
In Listing~\ref{lst:fipac:pac_check}, we show the corresponding assembly sequence.
We first insert an instruction sequence that patches the current CFI state to a valid PAC value using a constant for this location in the program.
Then, we use the \autiza instruction to verify the integrity of this PAC value.
On a control-flow deviation, applying the constant to the incorrect CFI state in \texttt{x28} generates an invalid PAC, which can be detected by the \autiza instruction.
If the check fails, the \autiza instruction traps, and the program is stopped.

CFI checks can be placed arbitrarily within the program.
\fipac supports three strategies: one check at the end of a program, a check at the end of every function, or a check at the end of every basic block in the program.
The check strategy has a direct impact on the performance and security of the design, which is discussed in Section~\ref{sec:fipac:evaluation}.

\paragraph{Key Management.}

ARM pointer authentication includes five key registers containing the generic key, the instruction, and data keys \texttt{A} and \texttt{B}.
By utilizing the \pacia instruction to compute the CFI state, \fipac uses the \texttt{APIAKey} for the keyed MAC.
The key is managed in kernel mode (\texttt{EL1}) and access to the key register is prohibited from user mode (\texttt{EL0})~\cite{armpac}.
To provide CFI protection with \fipac for the kernel as well, the key management, and therefore also the access permission, can be delegated to a higher privilege level (e.g. \texttt{EL2}).
As PA instructions do not differentiate between privilege levels, these instructions can use the PA keys in \texttt{EL0} and \texttt{EL1}.
To prevent cross-EL attacks~\cite{pacprojectzero}, \fipac protected user and kernel tasks can either use different keys for each privilege level (e.g. \texttt{APIAKey} for \texttt{EL0} and \texttt{APIBKey} for \texttt{EL1}), or the key manager in \texttt{EL2} could swap the keys on mode transitions.
As the key needs to be known at compile time by the toolchain to compute the CFI states, the prototype implementation of \fipac statically configures the \texttt{APIAKey} in a kernel module in \texttt{EL1}.
We discuss the dynamic configuration of the PA keys in Section~\ref{sec:fipac:futurew}.

\paragraph{Interrupts.}

\fipac natively supports interrupts and operating system interactions without any change.
When an interrupt diverts the control-flow to the kernel, it saves all registers of the user application including the current CFI state.
After resuming from the interrupt, the CFI state is restored, allowing the program to continue.

\subsection{Toolchain}
\label{sec:fipac:toolchain}

\begin{figure}[t]
  \begin{center}
    \setlength{\belowcaptionskip}{-5mm}
    \includegraphics[width=0.4\linewidth]{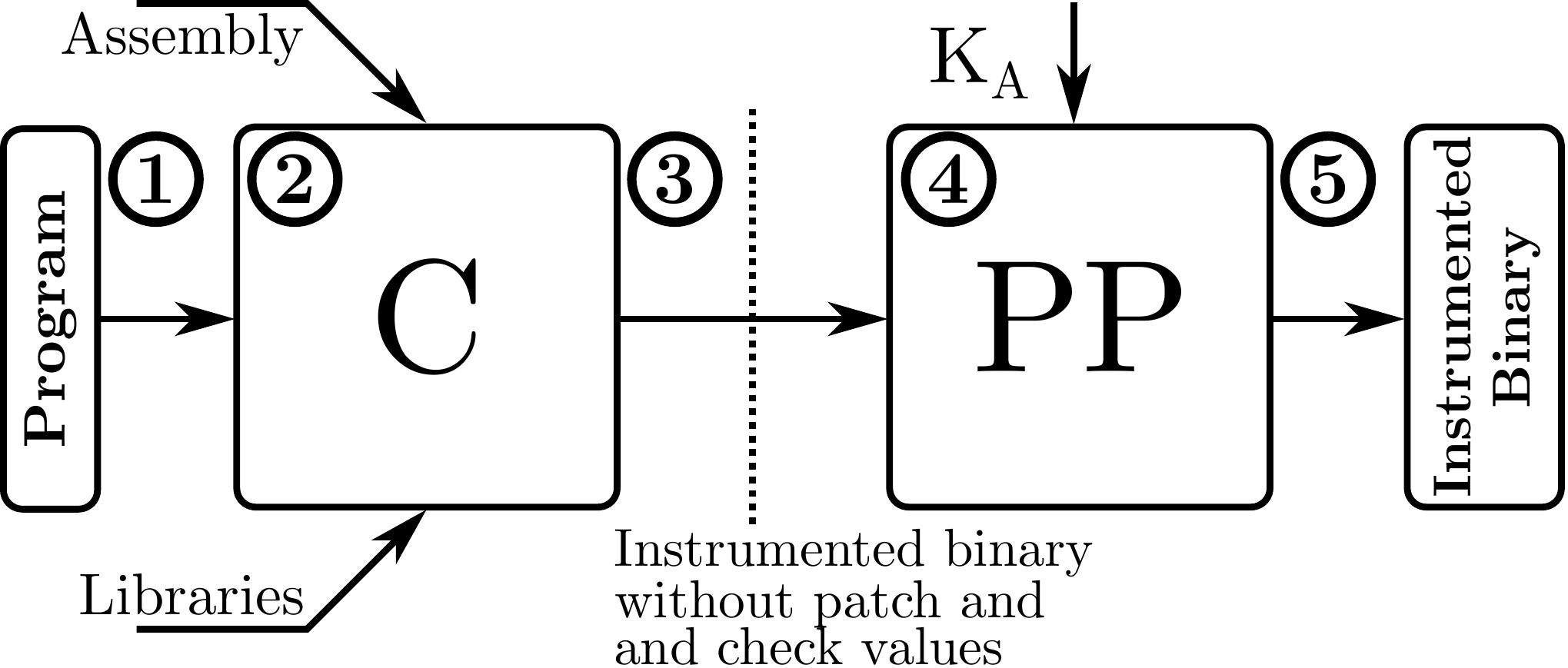}
    \caption{Custom toolchain to build protected binaries.}
    \label{fig:fipac:custom_toolchain}
  \end{center}
\end{figure}

Related work~\cite{coudray2015picon,DBLP:conf/sigmetrics/SchardlDDKLL18,DBLP:conf/cardis/WernerWM15,DBLP:conf/cardis/WernerWM15} either uses the compiler or dedicated post-processing tools operating on the binary, to perform the instrumentation.
Our prototype toolchain uses a combination of both approaches, which compilation flow we visualize in Figure~\ref{fig:fipac:custom_toolchain}.
We use a custom compiler~\circleds{2} based on the LLVM compiler framework, to insert all necessary state update and patch instructions during the compilation of a program~\circleds{1}.
The compiler emits an instrumented ELF binary~\circleds{3}, but the concrete state patches and check values are set to zero.
In a second step, we use a post-processing tool~(PP)~\circleds{4}, which has access to the whole compiled and linked binary.
This tool computes the CFI state throughout the program and replaces the update patches and check values with the concrete values.

Our toolchain fully supports instrumented or non-instrumented libraries, but only instrumented libraries have control-flow protection.
Instrumented libraries must be linked statically, such that the post-processing tool can replace the patch and check values for the compiled binary.
The toolchain also supports inline assembly and external assembler files.
However, it is the programmer's responsibility to insert the necessary state update and patch sequences to the assembly code.
In the second stage, our post-processing tool takes care of computing all expected states and inserting the patch updates.
If the assembly code is not instrumented, the code is still fully functional but does not have CFI protection.

\paragraph{Limitations.}

The toolchain currently only supports the instrumentation of programs written in the C programming language.
However, extending the support to other languages supported by LLVM compiler framework, \eg C++ or Rust, only requires more engineering work but no changes to the design of \fipac.

\subsubsection{LLVM Compiler}

To automatically compile binaries without user interaction, we develop a custom toolchain based on the LLVM compiler framework~\cite{DBLP:conf/cgo/LattnerA04}.
We extend the AArch64 backend and reserve the general purpose register \texttt{x28}, which is exclusively used to store the CFI state, disable tail calls, and ensure that functions have only a single return point.
For the CFI protection, we add two compiler passes providing the following instrumentation.

\paragraph{CFI Updates, Patches, and Direct Calls.}

At the beginning of each basic block, we insert the sequence for a PA-based CFI state update, as shown in Listing~\ref{lst:fipac:pac_state_update}.
This instruction sequence uniquely updates the CFI state for the current basic block based on the previous state value.
Furthermore, all direct calls are instrumented with state patches, which transform the current state to the beginning state of the called function.
Indirect calls are instrumented to stack the current CFI state and patch the CFI state for the intermediate indirect call signature.
After the indirect call, the pre-call state saved on the stack is retrieved and XORed to the CFI state to provide a link over the indirect call.

Moreover, this compiler instruments conditional branches and inserts the required patches for merging control-flows within a function.
To identify the locations of the needed CFI patches, we compute the inverted maximum spanning tree over the edges of the CFG of a function, defining the patch locations.

\begin{lstlisting}[float,language=c++,style=aarch64,mathescape=true,caption={Function entry points for indirect and direct calls.},label={lst:fipac:entrypoints},belowskip=-8mm]
  mov x1, #I_PATCH    ; Indirect call entry point
  eor x28, x28, x1    ; Patch to beginning state of function
  mov x1, #RET_PATCH  ; Load return patch
  b #8
  mov x1, #0          ; Direct Call entry point
  ...                 ; sets up zero patch
  eor x28, x28, x1    ; Apply return patch
  ret
\end{lstlisting}

\paragraph{Second Function Entry Point.}

Indirect calls require a more complicated instrumentation besides the instrumentation at the call site.
As discussed in Section~\ref{sec:fipac:design:icalls}, the function header of an indirectly called function needs to set up the patch value, which is used during the return of the function.
However, a function generally does not know how it was called and must support being called directly and indirectly.
Thus, every function requires two entry points, one for direct calls and a second one for indirect calls.

At the end of the backend pipeline of the compiler, we add a second machine function pass that adds a custom function entry point used for indirect calls, as shown in Listing~\ref{lst:fipac:entrypoints}.
In this entry point, we first patch the intermediate state of the indirect call to the beginning state of the called function (Line 1-3).
We then load, in Line~4, the CFI update patch, which is used during the return of the function, and then jump, in Line~5, over the direct call entry point.
When the function was called directly, it jumps to the direct call function entry point in Line~6, which sets up a zero-patch for the function return.
During the return of the function, Line~8 uses the previously configured return patch.
For direct calls, where the return patch in \texttt{x1} is zero, this statement has no effect, but for indirect calls, it patches the end state of the function to the intermediate return state for indirect calls.

The compiler, however, is not aware that the inserted instructions have control-flow and implement a second function entry point.
Thus, after compilation, direct calls also use the second entry point, which is exclusively for indirect calls.
We correct this during the post-processing stage, where all direct calls get rewritten to use the second entry point.

\subsubsection{Post-Processing Tool}
\label{sec:fipac:pp}

The post-processing tool is a custom tool, performing call rewriting, the correct CFI state computation, inserting the correct patch values, and computing the CFI check values.
It has access to the PA key and consumes the instrumented ELF binary, where all patches and checks are zero.
It first rewrites all direct calls to use the second function entry point (the first one is used for indirect calls exclusively).
Next, it recovers the control-flow of the whole program to compute to CFI state for every location in the program.
Every function is assigned a random starting signature, which is propagated through all PAC-based state updates of the function.
At a control-flow merge, the state values of both branches are known, thus, the post-processing tool can compute the justifying signature as the XOR-difference between both states.
It finally replaces the patch value \texttt{\#patch} with the previously computed justifying signature.
The post-processing tool knows the CFI state at every location in the program, thus it can also compute the XOR-differences to form a valid PAC.
For \autiza-based check sequences, it replaces \texttt{\#const} with the corresponding XOR-difference.

\section{Evaluation}
\label{sec:fipac:evaluation}

In this section, we discuss the security guarantees of \fipac and analyze  different security policies.
We further validate the correctness of our scheme by running the application-grade SPEC2017 benchmark and embedded benchmarks on a functional ARMv8.6-A model.
Finally, we describe our test setup to measure the runtime overhead for \fipac on the selected benchmarks and evaluate the overhead of different checking policies.

\subsection{Security Evaluation}
\label{sec:fipac:secevaluation}

\begin{figure}[t]
  \centering
  \includegraphics[width=0.45\linewidth]{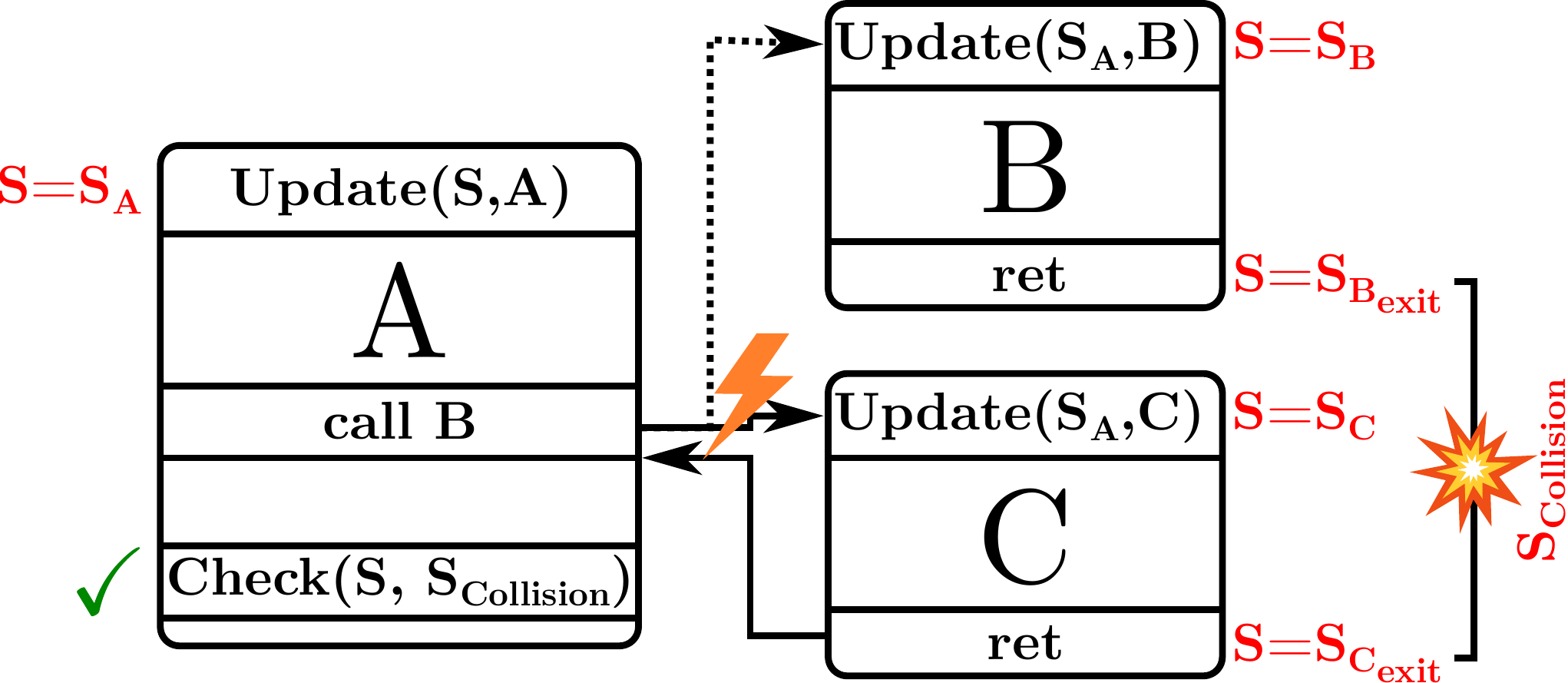}
  \setlength{\belowcaptionskip}{0mm}
  \caption{Redirection of control-flow from \texttt{B} to \texttt{C}. Due to a CFI state collision, the control-flow hijacking attack is not detected.}
  \label{fig:fipac:collision_prob}
\end{figure}

As highlighted in the threat model in Section~\ref{sec:fipac:threat}, \fipac considers a software and fault attacker aiming to hijack control-flow transfers between basic blocks.
To protect these control-flow transfers, \fipac performs a state update of the global CFI state $S$ at the end of every basic block.
This allows \fipac to detect inter basic block manipulations, \ie attacks on direct branches, direct calls, and indirect branches.
We provide real-world exploits in Appendix~\ref{sec:fipac:exploits} and show, how \fipac protects the programs from such attacks.

To prevent a software attacker from forging the CFI state, \fipac uses a keyed state update mechanism.
Equation~\ref{eq:fipac:state_update} depicts the state update function consisting of the secret key $K_A$, the current state $S$, and an unique identifier $Sig_{BB}$ for the basic block, \ie the program counter.
The secret key $K_A$, inaccessible by the adversary, is initialized at boot time and ensures that the attacker cannot forge a specific state value.

\begin{equation}
\begin{split}
 \label{eq:fipac:state_update}
  S &= \text{Update}(S, Sig_{BB}, K_A) \\
    &= S \oplus \text{MAC}_{K_A}\left( Sig_{BB} \right)_{PAC_{Size}}
\end{split}
\end{equation}
To efficiently conduct the state update, we utilize ARM's \pacia instruction.
Internally, \pacia uses the tweakable block cipher QARMA~\cite{DBLP:journals/tosc/Avanzi17} for computing the MAC.
Here, the state $S$ is used as the encryption input, the basic block identifier $Sig_{BB}$ as a tweak for the cipher, and the encryption output is used for the pointer authentication code.
Although the encryption output of QARMA is 64-bit in size, ARM pointer authentication limits the size of the PAC value to $PAC_{size}$ bits.
Since the PAC is stored in the upper unused bits of the pointer, the size of the virtual address space is the main limiting factor for the PAC size.
For the two typical Linux virtual address space configurations with 39- or 48-bit virtual addresses~\cite{aarch64va}, a maximum $PAC_{size}$ of 16- or 24-bits is possible~\cite{DBLP:conf/uss/LiljestrandNWPE19}.

\paragraph{CFI State Collision Probability.}

The size of the PAC $PAC_{size}$ and, therefore, the size of the state $State_{size}$ is essential for the security of \fipac.
Due to the truncation of the 64-bit output of QARMA to $PAC_{size}$, state collisions are possible with a probability of $P_{Collision} = \frac{1}{2^{PAC\_SIZE}}$, which can lead to a bypass of the CFI protection.
Figure~\ref{fig:fipac:collision_prob} illustrates a control-flow hijack, redirecting the call from \texttt{B} to \texttt{C} by using a software vulnerability or a fault attack.
When returning to the caller \texttt{A}, the state mismatch $S_{C_{exit}} \neq S_{B_{exit}}$ should be detected by the checking mechanism of \fipac.
However, with probability $P_{Collision}$, a state collision $S_{C_{exit}} = S_{B_{exit}} = S_{Collision}$ occurs, and the control-flow attack remains undetected.

\paragraph{Checking Policy.}

\begin{figure}[t]
  \centering
  \includegraphics[width=0.6\linewidth]{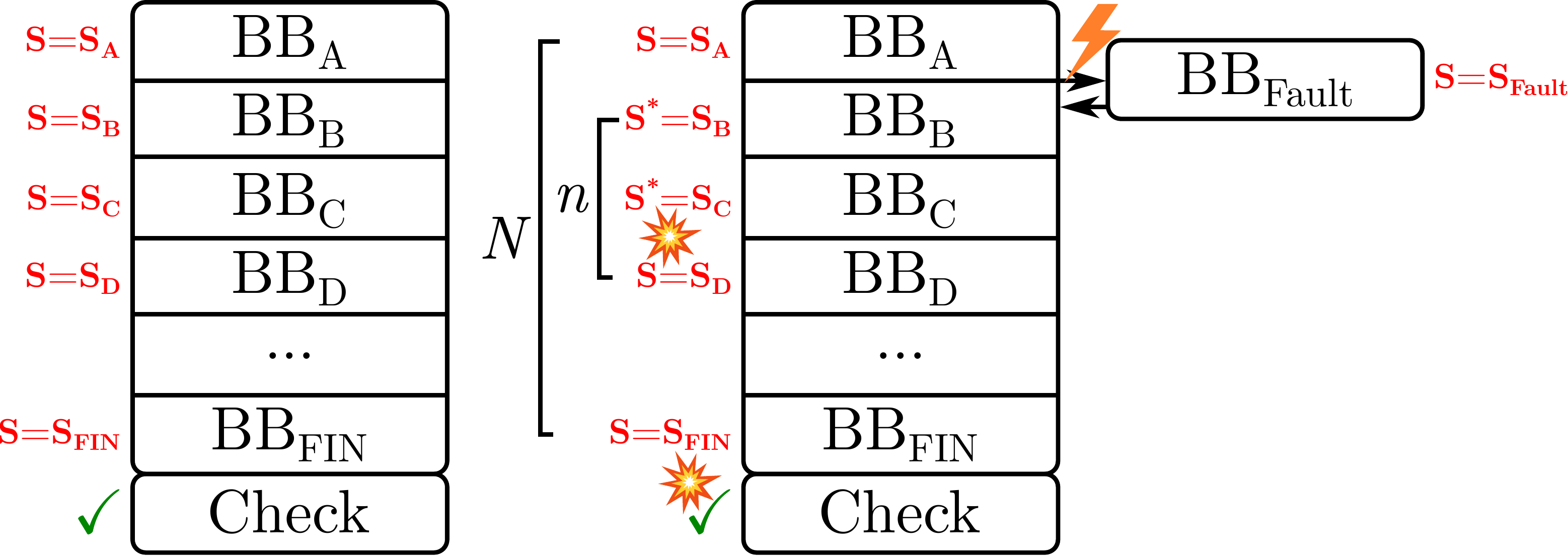}
  \caption{Example of a too coarse-grained checking policy. After $n$ state updates, a collision rectifies the faulty state $S^*$ to $S$.}
  \label{fig:fipac:collision_min_prob}
\end{figure}

To reliably detect state collisions, the sufficient placement of CFI state checks, \ie the checking policy, is crucial for the security of \fipac.
However, properly placing CFI checks within the program is a challenging problem with no general solution.
Figure~\ref{fig:fipac:collision_min_prob} illustrates the problem of a too coarse-grained checking policy.
On the left side, a valid control-flow from basic block $BB_A$ to $BB_{FIN}$ is shown.
On the right side of the figure, the attacker manages to redirect the control-flow to $BB_{Fault}$ and therefore alter all subsequent states to $S^*$.
However, with a probability of $P_{Collision}$, a state collision occurs after each state update.
In this example, after $n$ updates, a collision occurred and $S^*$ becomes $S_D$.
Thus, the state $S$ is valid again and the control-flow hijack cannot be detected anymore in further CFI checks.

\begin{figure}[t]
  \begin{minipage}[t]{.44\textwidth}
    \begin{figure}[H]
      \centering
      \setlength{\belowcaptionskip}{-5mm}
      \includegraphics[width=1\linewidth]{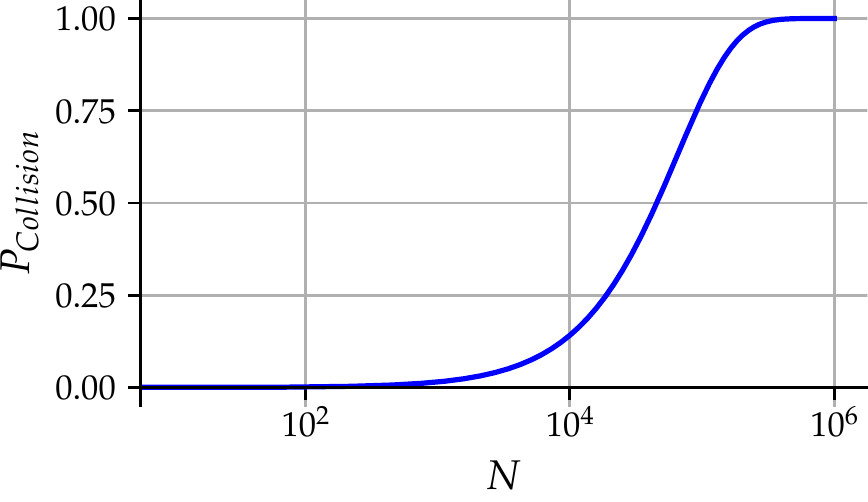}
      \caption{Collision probability after $N$ state updates.}
      \label{fig:fipac:minimum_probability}
    \end{figure}
  \end{minipage}\hfill
  \begin{minipage}[t]{.44\textwidth}
    \begin{figure}[H]
      \centering
      \setlength{\belowcaptionskip}{-5mm}
      \includegraphics[width=1\linewidth]{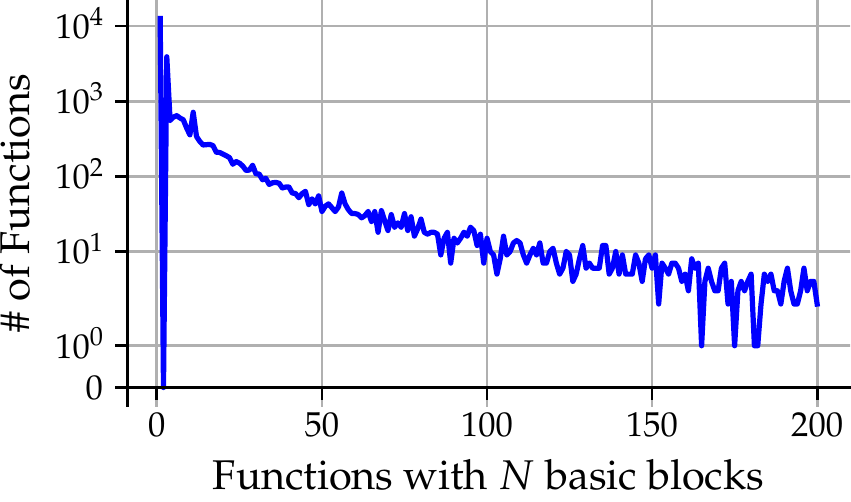}
      \caption{Number of functions with $N$ basic blocks.}
      \label{fig:fipac:nr_function}
    \end{figure}
  \end{minipage}
\end{figure}

In order to give a quantitative measure on the security of the placement of checks, we analyze the probability that undetectable state collisions occur between subsequent CFI checks.
Equation~\ref{eq:fipac:minimum_probability} shows the minimum probability that a state collision occurs in one of $N$ state updates.
As illustrated in Figure~\ref{fig:fipac:minimum_probability}, after the execution of 100,000 state updates, a state collision probability of 78\,\% is given.
With the execution of 500,000 state updates, and therefore the same number of basic blocks, a collision occurs with almost 100\,\%.

\begin{equation}
 \label{eq:fipac:minimum_probability}
   \text{MP}_{Collision_N}=1-\left(1-\frac{1}{2^{PAC\_SIZE}}\right)^N
\end{equation}

Selecting the checking policy is a trade-off between security and performance.
Although the most precise policy, \ie a check at each basic block, maximizes the probability of detecting a control-flow hijacking attack, the performance overhead also increases.
While a loose checking policy, \eg a check at the end of the program, might be sufficient for small programs, larger programs with a high number of executed basic blocks might be vulnerable to attacks.
Between these two policies, arbitrary other checking strategies can be selected; for example, a check at the end of each function.
A more advanced placement strategy of checks can incorporate additional information, \eg runtime profiling.
This allows the compiler to better decide where checks are needed to enforce a lower bound of the minimum detection probability of CFI errors.

However, we argue that a CFI check at the end of a function is a good trade-off between runtime overhead and security.
For example, the SPEC2017 benchmarks consists of 28391 functions.
Interestingly, 12583 of these functions, or 44\,\%, contain only a single basic block with a CFI check at the end.
Thus, calling such a function is equivalent with performing a CFI state check at call site.
For example, calling this function within a loop containing no explicit checks, implicitly performs a CFI state validation at each loop iteration.

Figure~\ref{fig:fipac:nr_function} depicts the occurrence of functions with a certain number of basic blocks.
The number of functions with a small number of basic blocks is much larger than functions comprising a large number of basic blocks.
Almost 75\,\% of all functions consists of less than 13 basic blocks with a CFI check at their end.
This property is in favor with our checking policy, since smaller functions perform a CFI validation earlier than large functions.
Thus, the detection probability of a CFI state mismatch is higher.
Summarizing, we expect that a CFI check at the end of each function is a good trade-off for a static checking policy.

\paragraph{Skipping an \autiza Instruction.}

\fipac does not detect instruction skips \emph{within} a basic block as there is no intra basic block protection.
A control-flow hijack combined with an instruction skip over following check instruction is not detectable.
However, depending on the checking policy, a new check instruction might occur at the end of the next basic block or at the end of the function.
Since a this point of the attack, the CFI state is invalid, it requires the attacker to also skip \emph{all} subsequent check instructions such that the control-flow attack is not detectable.

The same mechanism also defeats a classical ROP attack.
When performing a ROP attack and jumping to the ROP gadget, a check instruction might occur at the end of the basic block or function.
It requires to skip \emph{all} subsequent check instructions for every ROP gadget in the ROP chain such that the attack is not detectable.

\subsection{Security Comparison}
\label{sec:fipac:secomparison}

Previous CFI schemes either consider a software attacker~\cite{DBLP:journals/tissec/AbadiBEL09, DBLP:conf/uss/TiceRCCELP14, DBLP:conf/uss/ZhangS13, DBLP:conf/sp/EvansFGOTSSRO15, DBLP:conf/ccs/MashtizadehBBM15, DBLP:conf/uss/LiljestrandNWPE19, DBLP:journals/corr/abs-1905-10242} or a fault attacker~\cite{DBLP:journals/tr/OhSM02a, DBLP:conf/cgo/ReisCVRA05, DBLP:conf/iolts/VenkatasubramanianHM03, DBLP:conf/esorics/LalandeHB14, DBLP:journals/compsec/HeydemannLB19}, but they do not cover both attackers in their threat model.
Software CFI schemes, like PARTS~\cite{DBLP:conf/uss/LiljestrandNWPE19}, CPI~\cite{DBLP:conf/sp/EvansFGOTSSRO15}, or CCFI~\cite{DBLP:conf/ccs/MashtizadehBBM15}, enforce control-flow integrity at a coarse granularity by protecting a wide range of forward- and backward edges, \eg code-pointers and return addresses, on function level.
Although these approaches mitigate a software attacker exploiting a memory vulnerability, they fail to protect against a fault attacker.
To protect the control-flow from fault attacks, fault CFI schemes enforce the CFI at a much finer granularity, \ie on basic block or instruction level.
In contrast to a pure software attacker exploiting memory vulnerabilities, a precise fault can tamper with direct and indirect control-flow transfers.
While software-based fault CFI schemes protect all control-flow transfers from faults, they fail to protect the program from a software adversary attacking the state update function using a memory vulnerability.

To protect the system against control-flow attacks from a fault and software attacker, it is tempting to naïvely combine existing schemes such as PARTS with a fault CFI protection mechanism, \eg CFCSS~\cite{DBLP:journals/tr/OhSM02a}.
While these schemes are secure in their own threat model, a combined fault and software attack could bypass them.
First, the adversary gains control over a register used to update the FCFI state.
Then, the attacker redirects the control-flow to an illegitimate function, \eg with a fault.
Finally, when executing this function, the tampered register is used for the state update, thus, can forge a valid CFI state.

To protect against fault and software attacks and to support a large-scale deployment, \fipac fulfills the key requirements stated in Section~\ref{sec:fipac:sfcfi}.
First, \fipac comprehensively enforces the control-flow integrity for transfers between basic blocks.
Hence, our scheme operates on a much finer granularity than typical software CFI schemes.
Second, \fipac uses, in comparison to fault CFI schemes, a keyed state update function to mitigate attacks targeting to manipulate the global CFI state.
Furthermore, \fipac is implemented in software and is applied automatically during compilation via a custom toolchain.
\enlargethispage{3mm}
\subsection{Functional Evaluation}

To evaluate the functional correctness of \fipac, we compiled the SPEC2017~\cite{Spec2017}, Embench~\cite{embench-iot}, and CoreMark~\cite{coremark} benchmark with our custom LLVM-based toolchain and executed these instrumented binaries on the QEMU processor emulator~\cite{qemu}.
Since the latest QEMU release, version 6.0, only implements the ARMv8.5-A architecture, and thus does not support \textit{EnhancedPAC2} and \textit{FPAC}, we extended the QEMU codebase to support this feature.
In QEMU, we started the 5.4.58 Linux kernel and initialized the PA keys during the boot procedure before starting the benchmarks.

\subsection{Performance Evaluation}

\fipac exploits ARM pointer authentication of ARMv8.6-A.
However, to the best of our knowledge, there is currently no publicly available device or development board available supporting this architecture.
Instead, to conduct our performance evaluation on real hardware, we use the Raspberry Pi 4 Model B~\cite{rpi4}, and run our experiments on the 64-bit Raspberry Pi OS~\cite{raspbian}.
Since the ARM Cortex-A72 CPU is based on ARMv8-A, and does not support pointer authentication, we emulate the runtime overhead of the PA instructions in software.
We replace all PA instructions with their PA-analogue, \ie four consecutive XORs.
PARTS~\cite{DBLP:conf/uss/LiljestrandNWPE19} introduced and evaluated this sequence of operations to model the timing behavior of native PA instructions, which is also used in related work~\cite{DBLP:journals/corr/abs-1905-10242}.


\subsubsection{SPEC 2017}

To measure the performance overhead of \fipac on applications, we compiled all C-based benchmarks with OpenMP support disabled of the SPECspeed 2017 Integer test suite with our LLVM-based toolchain.
Furthermore, we enabled three different checking policies, from coarse-grained to fine-grained checks, to compare the performance penalty introduced by them.
More concretely, we configured \fipac to insert a CFI check at the end of the program, at the end of every function, or at each basic block.

In Table~\ref{tab:fipac:codesize_spec}, we summarize the code overhead with the three different checking policies. 
As expected, the checking policy with a single check at the end of the program has the lowest overhead.
Verifying the CFI state at the end of every basic block has the largest penalty as it requires three additional instructions per basic block.
Interestingly, placing a CFI check at the end of every function has only a small impact on the code size compared to a single CFI check at the end of the program.
Due to this small increase in code size but its stronger security guarantees, this checking policy is a good trade-off.
{
\vspace{0.05cm}
\begin{minipage}[b][][b]{0.45\textwidth}
  \centering
  \includegraphics[width=1\linewidth]{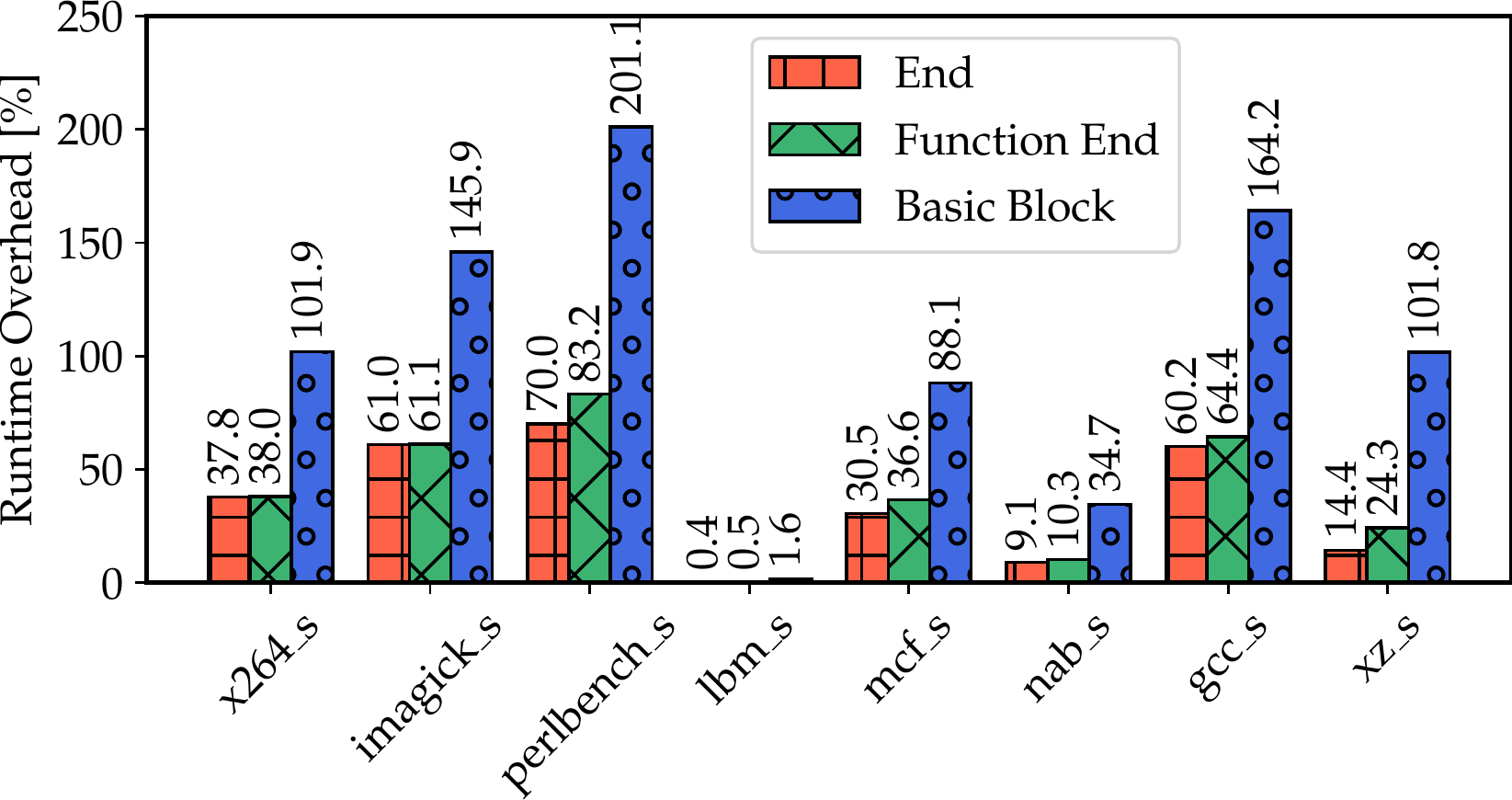}
  \captionof{figure}{\label{fig:fipac:runtime_spec}Runtime overhead for SPECspeed 2017.}
\end{minipage}\hfill
\begin{minipage}[b][][b]{0.45\textwidth}
  \centering
  \small
  \begin{tabular}{llll}
      \toprule
      \textbf{Testcase} & \textbf{End}  & \textbf{Fend} & \textbf{BB} \\
      \midrule
      lbm\_s           & 19.19 & 20.85 & 33.30\\
      sgcc             & 74.31 & 77.62 & 130.71\\
      xz\_s            & 48.08 & 50.65 & 85.78\\
      perlbench\_s     & 66.00 & 67.22 & 120.18\\
      nab\_s           & 63.96 & 65.38 & 117.15\\
      x264\_s          & 41.56 & 42.60 & 73.40\\
      imagick\_s       & 58.44 & 59.99 & 101.91\\
      mcf\_s           & 61.29 & 62.78 & 115.92\\
      \midrule
      \textbf{Average} & \textbf{54.10} & \textbf{55.89} & \textbf{97.29}\\
      \hline
    \end{tabular}
  \captionof{table}{\label{tab:fipac:codesize_spec}Code size overhead for SPECspeed 2017.}
\end{minipage}
}
\\Figure~\ref{fig:fipac:runtime_spec} depicts the runtime overhead of \fipac compared to the baseline without any protection.
As expected, the coarse-grained checking policy with a single check at the end of the program introduces the smallest runtime overhead of 35\,\% on average.
Consequently, the fine-grained checking policy with CFI checks at the end of every basic block has the largest runtime penalty of 105\,\% on average.
Interestingly, the intermediate checking policy with a check at the end of each function introduces a runtime overhead of 39\,\% on average.
This is only a small increase compared to a single check at the end, but provides much better security guarantees.
Note, since the control-flow of the \texttt{lbm\_s} benchmark is mostly linear and this test performs a large number of expensive floating point operations, the impact of \fipac is rather small.
The runtime overhead for difference security polices is between \specruntimerange outperforming related work with overheads between 107--426\,\%~\cite{DBLP:conf/dft/GoloubevaRRV03}.

\subsubsection{Embench}

\begin{figure}[t]
  \begin{center}
    \setlength{\belowcaptionskip}{-5mm}
    \includegraphics[width=1\linewidth]{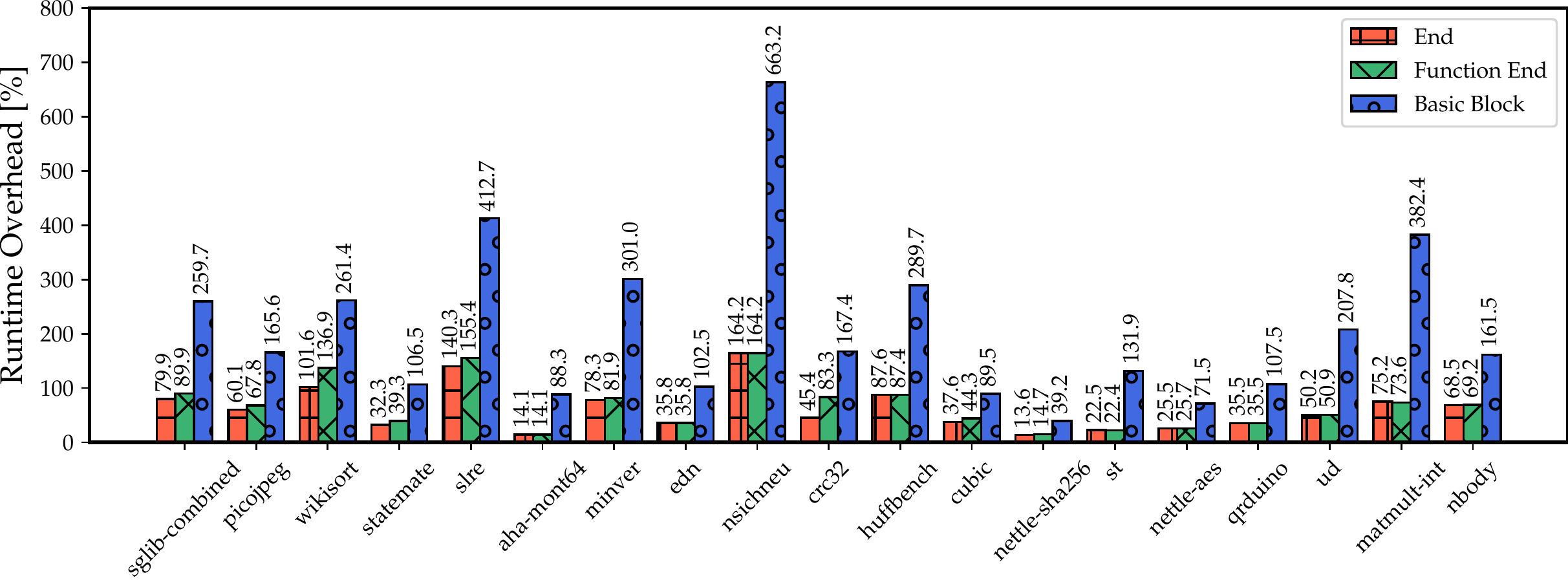}
    \caption{Runtime overhead for Embench.}
    \label{fig:fipac:runtime_embench}
  \end{center}
\end{figure}

To evaluate \fipac on embedded workloads, we use Embench, largely derived from the BEEBS benchmark suite~\cite{beebs}.
The code overheads, as depicted in Table~\ref{tab:fipac:codesize_embench} in Appendix~\ref{sec:fipac:codeembench}, are between 57.8\,\% and 110.5\,\%, depending on the selected checking policy.
On average, the runtime overheads, as shown in Figure~\ref{fig:fipac:runtime_embench}, of Embench are between 61.4\,\% and 211.0\,\%, which are slightly higher compared to SPEC.
We observed this increased overhead due to the fact that Embench's codebase is small with a larger number of control-flow transfers compared to application-grade benchmarks such as SPEC2017.

\section{Discussion}
\label{sec:fipac:futurew}

In this section, we discuss the hardware requirements of \fipac, how it can be implemented on other architectures, and future improvements of the countermeasure.

\paragraph{\fipac Hardware Requirements.}

\fipac requires pointer authentication with EnhancedPAC2 and FPAC of ARMv8.6-A, to maintain the global CFI state.
At the time of writing, there is no open hardware available implementing ARMv8.6-A yet.
However, ARM presented the ARM Cortex-A78C~\cite{armcortexa78c} last year and the announced Qualcomm Snapdragon SC8280XP probably implements this micro-architecture.
Although ARMv8.6-A is not yet widely available in existing processors, ARM already announced the successor instruction set ARMv9-A~\cite{armv9} with support for EnhancedPAC2 and FPAC.
Hence, we expect that new designs, \eg Apple's upcoming processors, to feature ARMv8.6-A or even ARMv9-A.

\paragraph{\fipac on ARMv8.3-A.}

Although \fipac is designed for ARMv8.6-A, it also could be implemented on ARMv8.3-A architectures with the following adaptions.
ARMv8.3-A PA instructions only compute a new PAC without accumulating it.
Thus, the state accumulation must be done manually using an additional \texttt{eor} instruction per state update.
This increases the overhead of a CFI state update to 3 instructions and requires one more temporary register to hold the PAC.
Furthermore, the \texttt{autiza} instruction in ARMv8.3-A cannot be used to implement a check as it does not trap on an incorrect verification.
However, ARMv8.3-A features the \texttt{blraa} instruction, a branch with link to register operation with pointer authentication, which traps if the jump-target address contains an invalid PAC.
This instruction can be misused to perform a CFI check for \fipac.
Similar to the original design of \fipac, we transform the known CFI state to a valid PAC with the address being the program location of the next instruction.
When executing this branch, it first verifies the target address and, if valid, unconditionally jumps to the next instruction.
If the PAC, and therefore also the CFI state, is invalid, the verification mechanism traps and stops the program.
Both solutions increase the code size overhead but also the runtime overhead compared to the prototype implementation of \fipac based on ARMv8.6-A.

Similar to ARMv8.6-A, currently, there is no open hardware available implementing ARMv8.3-A yet.
Although Apple already offers cores, such as the M1 and A14~\cite{applesocsecurity}, supporting PA of ARMv8.3-A, Apple restricts the usage of this feature.
iOS applications are not allowed to load custom kernel modules, thus \fipac cannot configure the PA keys in the kernel.
\fipac may run on the Apple M1 core with PA of ARMv8.3-A.
However, we currently do not have access to such a device and it requires future research to clarify if setting PA keys is possible in the EL1 kernel mode or if Apple restricts it.

\paragraph{\fipac on Other Architectures.}

The design of \fipac is generic, and thus, could also be implemented on other architectures with similar primitives.
It is tempting to implement \fipac on x86 with the AES-NI~\cite{aesni} extension, which supports partial encryption, \ie one round, with a single instruction.
However, we see limitations with this approach.
First, such an operation operates on a 128-bit state, also requiring to embed 128-bit patch values for control-flow transfers in the binary.
Second, a single AES-NI operation only computes one round and therefore only provides scrambling and no cryptographic strength.
Third, it requires the encryption keys to be held in general purpose registers of the processor.
Thus, there is no isolation of keys between the user application and the kernel or to higher privilege levels.
Summing up, we do not envision \fipac to be implemented with AES-NI.

\paragraph{Dynamic Key Handling.}

The prototype of \fipac uses a static PA key, which is configured by the OS.
However, ARM pointer authentication supports up to five encryption keys for different domains.
By using different keys for \fipac, the CFI protection between different domains, \eg the kernel and user programs, can be isolated.

For better isolation between applications, a future version of \fipac could embed the PA key in the binary, thus allowing different applications to use different PA keys.
Existing key exchange algorithms are then used to protect the embedded PA key.
The OS, which starts a binary and has access to a private key for the key exchange, can read the PA key and configure the system before starting the protected binary.

The previous approaches use a static key per binary, such that all executions use the same PA key.
To dynamically change the PA key, the post-processing step could be integrated to the OS.
Before starting the application, the OS chooses a random PA key and performs the post-processing step, \ie the computation of the CFI states, patches, and check values on demand.
Thus, every invocation of the application is different in terms of \fipac related patch and check values, which also hardens the attack surface.


\paragraph{Instruction Granular Protection.}

\fipac does not protect the linear instruction sequence within a basic block.
If a more fine granular protection in a refined threat model is required, \ie intra basic block security, \fipac further supports the placement of state updates within security-critical basic blocks.
For example, a CFI state update can be placed after every instruction, to emulate an instruction-granular CFI protection.
Based on the state update of Listing~\ref{lst:fipac:pac_state_update}, it requires two additional operations to protect one instruction.
However, automatically identifying such critical pieces of code is a challenging task and not in the scope of this work.
Instead, it requires the developer to manually place a check, \eg via inline assembly.


\paragraph{Compatibilty.}

\fipac currently uses the instruction address for the signature computation.
When ASLR is enabled, it leads to randomized signatures not being compatible with the static computation.
This problem can be solved by using static numbers for computing the signatures or by integrating dynamic key handling in the operating system.

\fipac is a software-based CFI protection scheme, and therefore, comes with certain degrees of flexibility, compared to hardware-centric approaches.
The design of \fipac is flexible and supports arbitrary checking policies on the same system.
For example, a critical application, \eg running within a trusted environment or an enclave, can have a stronger checking policy than a non-critical user application.
Furthermore, \fipac is backward compatible and supports non-instrumented applications without a change.

\section{Conclusion}
\label{sec:fipac:conclusion}

With the rise of new attack methodologies, fault and software attacks are omnipresent on all commodity devices and require protection.
While there exist different defenses, they are not sufficient when considering both fault and software attacks.

In this work, we presented \fipac, a fine-granular software-based CFI protection scheme for upcoming ARM-based hardware.
\fipac offers fine granular control-flow protection on basic block level for both fault and software attacks.
The design exploits a cryptographically secure state update function, which cannot be recomputed without knowing a secret key.
\fipac utilizes ARM pointer authentication orf ARMv8.6-A, to efficiently implement the keyed CFI state update and checking mechanism.
Furthermore, we provide a custom toolchain to automatically instrument and protect applications without user interaction.
The evaluation of \fipac with the SPEC2017 benchmark with different security policies shows an average runtime overhead between \specruntimerange and is slightly larger for small embedded benchmarks.
\fipac is an efficient software-based CFI protection, requires no hardware changes, and outperforms related work.

\ifanonymous
\else
\section*{Acknowledgments}
This project has received funding from the European Research Council (ERC) under the European Union’s Horizon 2020 research and innovation programme (grant agreement No 681402).
\fi

\bibliographystyle{plain}
\bibliography{bibliography}

\newpage

\appendix


\section{Code size overhead for Embench}
\label{sec:fipac:codeembench}

In Table~\ref{tab:fipac:codesize_embench}, we show the detailed code overhead for all benchmarks of Embench for the three checking policies of \fipac.
\begin{table}[bh]
  \centering
  \small
  \singlespacing
  \caption{Code size overhead for Embench.}
  \label{tab:fipac:codesize_embench}
  \begin{tabular}{llll}
\toprule
\textbf{Testcase} & \textbf{End [\%]}  & \textbf{Function End [\%]} & \textbf{Basic Block [\%]} \\
\midrule
huffbench        & 70.04 & 76.03 & 119.19\\
slre             & 81.72 & 86.37 & 146.78\\
cubic            & 31.43 & 34.04 & 56.48\\
nbody            & 70.19 & 77.39 & 116.09\\
minver           & 74.24 & 80.79 & 126.64\\
sglib-combined   & 81.24 & 86.47 & 147.62\\
st               & 69.74 & 76.49 & 114.74\\
matmult-int      & 39.50 & 43.49 & 64.18\\
statemate        & 61.68 & 65.08 & 112.24\\
crc32            & 48.16 & 53.97 & 76.17\\
aha-mont64       & 70.82 & 77.66 & 117.22\\
qrduino          & 53.38 & 56.27 & 99.46\\
ud               & 71.66 & 78.07 & 119.44\\
edn              & 43.78 & 47.93 & 71.36\\
nettle-sha256    & 46.07 & 50.65 & 81.62\\
wikisort         & 60.25 & 64.86 & 105.86\\
nettle-aes       & 19.59 & 21.46 & 32.95\\
picojpeg         & 50.73 & 53.49 & 88.24\\
nsichneu         & 54.07 & 55.27 & 113.76\\
\midrule
\textbf{Average} & \textbf{57.81} & \textbf{62.41} & \textbf{100.53}\\
\bottomrule
\end{tabular}
\end{table}

\section{Example Exploits}
\label{sec:fipac:exploits}

This section shows two example control-flow exploits and how \fipac protects against.

\paragraph{NaCl Sandbox Escape via Rowhammer.}

In~\cite{rowhammerexploit}, Rowhammer is used to perform a sandbox escape exploit out from NaCl.
The Rowhammer effect is used to manipulate a jump instruction in the memory to an attacker-controlled destination.
In Listing~\ref{lst:fipac:nacl}, the final jump instruction is faulted to use the attacker-controlled register \texttt{ecx} as the jump target.
\begin{lstlisting}[mathescape = true,basicstyle = \small,language=c++,style=aarch64,mathescape=true,emph={andl,addq,jmp,eax,rax},emphstyle=\color{red},caption={NaCl attack gadget.},label={lst:fipac:nacl}]
  andl $\textdollar$~31, %eax  // Truncate and align to 32 bits.
  addq %r15, %rax  // Add %r15, the sandbox base address.
  jmp *%rax        // Indirect jump.
\end{lstlisting}
After jumping to the attacker-defined code position, \fipac detects the control-flow manipulation at the subsequent check instruction.
As discussed in the design of \fipac, fine-grained checking policies, such as the end of every basic block or at the end of every function, are supported.
When executing such a check instruction, \fipac detects the control-flow hijack since the CFI state is invalid.

\paragraph{RCE on an Electronic Control Unit.}

In~\cite{nasahl2019attacking}, a combined software and fault attack was used to gain remote code execution~(RCE) on an automotive electronic control unit~(ECU).
\begin{lstlisting}[language=c++,style=aarch64,mathescape=true,emph={pc,r1},emphstyle=\color{red},caption={Instruction corruption to load \texttt{r2} into \texttt{pc}.},label={lst:fipac:ldr}]
  ldr r1, [r2, #4]
  ldr pc, [r2, #4]
\end{lstlisting}
Here, as depicted in Listing~\ref{lst:fipac:ldr}, a targeted fault was used to corrupt a load instruction to modify the program counter.
As register \texttt{r2} either is directly attacker controllable, or can be modified using a memory vulnerability, the attacker can redirect the control-flow to an arbitrary code position.
While the concrete attack is not directly applicable to AArch64, as the program counter cannot be manipulated via a load instruction, \cite{timmersfault} highlights alternative methods for corrupting instructions to redirect the control-flow. 
Independently of the actual targeted instruction to corrupt, a fault is used to modify the program counter in an attacker-controlled way, yielding an arbitrary jump primitive.
\fipac is able to protect the program from such control-flow hijacks, as the global CFI state does not match the expected CFI state determined at compile time for this position in the code.
Hence, at the next \fipac check instruction, \eg at the end of the function the attacker redirected the control-flow to, \autiza traps and \fipac detects the attack.

\end{document}